\documentclass[a4paper]{article}
\usepackage{epsfig}

\usepackage[round,colon]{natbib}
\renewcommand{\cite}{\citep}
\title{Learning curves for Gaussian process regression: Approximations
and bounds}

\author{Peter Sollich, Anason Halees\thanks{Current address: Biostatistics, Epidemiology and Scientific Computing,
King Faisal Specialist Hospital and Research Center, MBC\#03, P.O.Box
3354, Riyadh 11211, Saudi Arabia. Email: ahalees@kfshrc.edu.sa}\\[3mm]
Department of Mathematics, King's College London\\
Strand, London WC2R 2LS, U.K. Email: peter.sollich@kcl.ac.uk
}

\date{}

\newcommand{\teach}{\theta^*}
\newcommand{\stud}{\theta}
\newcommand{\meanstud}{\hat{\theta}}
\newcommand{\noise}{\sigma^2}
\newcommand{\noiseinv}{\sigma^{-2}}
\newcommand{\KK}{{\mathbf K}}
\newcommand{\kk}{{\mathbf k}}

\newcommand{\yy}{{\mathbf y}}
\newcommand{\T}{^{\mathrm T}}
\newcommand{\tr}{{\rm tr\,}}
\newcommand{\eps}{\epsilon}

\newcommand{\epsLC}{\epsilon_{\rm LC}}
\newcommand{\epsLCo}{\epsilon_{\rm LC1}}
\newcommand{\epsUC}{\epsilon_{\rm UC}}

\newcommand{\nUC}{n'}

\newcommand{\epsUO}{\epsilon_{\rm UO}}
\newcommand{\epsWV}{\epsilon_{\rm WV}}
\newcommand{\epsMW}{\epsilon_{\rm MW}}
\newcommand{\epsOV}{\epsilon_{\rm OV}}
\newcommand{\epsTWO}{\epsilon_{\rm TWO}}
\newcommand{\epsPl}{\epsilon_{\rm Pl}}
\newcommand{\epsemp}{\epsilon_{\rm emp}}

\newcommand{\lam}{\lambda}
\newcommand{\xl}{x_l}
\newcommand{\yl}{y_l}
\newcommand{\xm}{x_m}
\newcommand{\ym}{y_m}
\newcommand{\LL}{{\mathbf \Lambda}}
\newcommand{\Ph}{{\mathbf \Psi}}
\newcommand{\mident}{{\mathbf I}}
\newcommand{\GG}{{\mathbf G}}
\newcommand{\Gfluc}{{\mathcal G}}
\newcommand{\new}{\mbox{\boldmath $\psi$}}

\newcommand{\Phii}{{\mathbf \Phi}}
\newcommand{\ma}{{\bf A}}
\newcommand{\mU}{{\bf U}}
\newcommand{\mv}{{\bf V}}

\newcommand{\bi}{\begin{itemize}}
\newcommand{\ei}{\end{itemize}}

\newcommand{\be}{\begin{equation}}
\newcommand{\ee}{\end{equation}}

\newcommand{\bea}{\begin{eqnarray}}
\newcommand{\eea}{\end{eqnarray}}
\newcommand{\beastar}{\begin{eqnarray*}}
\newcommand{\eeastar}{\end{eqnarray*}}

\newcommand{\lav}{\left\langle}
\newcommand{\rav}{\right\rangle}

\newcommand{\half}{\frac{1}{2}}

\newcommand{\eq}[1]{~(\ref{#1})}
\newcommand{\eqq}[2]{~(\ref{#1},\ref{#2})}

\newcommand{\order}{{{\mathcal O}}}

\newcommand{\ie}{{\it i.e.}}
\newcommand{\eg}{{\it e.g.}}

\def\(#1){(\ref{#1})}

\newcommand{\colvec}[2]{\left(\!\!\begin{array}{c}#1\\#2\end{array}\!\!\right)}

\newcommand{\al}{\alpha}
\newcommand{\bet}{\beta}
\newcommand{\gam}{\gamma}
\newcommand{\ca}{c_\al}
\newcommand{\na}{n_\al}
\newcommand{\xa}{x^\al}
\newcommand{\xb}{x^\bet}
\newcommand{\xg}{x^\gam}
\newcommand{\pa}{p_\al}
\newcommand{\pb}{p_\bet}
\newcommand{\pg}{p_\gam}

\newcommand{\CC}{{\mathbf C}}
\newcommand{\Ct}{{\mathbf {\tilde {C}}}}
\newcommand{\phiv}{\mbox{\boldmath $\phi$}}
\newcommand{\thv}{\mbox{\boldmath $\theta$}}
\newcommand{\Lv}{{\mathbf L}}
\newcommand{\Cov}{{\mathbf V}}

\newcommand{\MM}{{\mathbf M}}
\newcommand{\PP}{{\mathbf P}}
\newcommand{\QQ}{{\mathbf Q}}
\newcommand{\OO}{{\mathbf O}}

\newcommand{\ev}{{\mathbf e}}
\newcommand{\etam}{\mbox{\boldmath $\eta$}}

\begin{document}

\maketitle

\begin{abstract}
We consider the problem of calculating learning curves (i.e., average
generalization performance) of Gaussian processes used for
regression. On the basis of a simple expression for the generalization
error, in terms of the eigenvalue decomposition of the covariance
function, we derive a number of approximation schemes. We identify
where these become exact, and compare with existing bounds on learning
curves; the new approximations, which can be used for any input space
dimension, generally get substantially closer to the truth. We also
study possible improvements to our approximations. Finally, we use a
simple exactly solvable learning scenario to show that there are
limits of principle on the quality of approximations and bounds
expressible solely in terms of the eigenvalue spectrum of the
covariance function.
\end{abstract}

\iffalse
\section{Outline}

\bi
\item General explanation about GPs, include 3d demo plots
\item Explain approximations; char curves for LC into appendix
\item Improving the approx's (also motivated by search for tighter
bounds - mention one-step bound for LD here, calc'n in appendix):
discrete n correction (char curve bit in appendix too), in principle
correction for decoupling but get new order parameters
\item Many approx's depend only on eigenvalue spectrum; true for *all*
if stationary covariance function. How good can these be? Degenerate
example: Eigenvalues are not enough, even for stationary cov fn. No
nontrivial eigenvalue-dependent bound for general cov fn.
\ei
\fi

\section{Introduction: Gaussian processes}

Within the neural networks community, there has in the last few years
been a good deal of excitement about the use of Gaussian processes as
an alternative to feedforward networks (see
\eg~\cite{WilRas96,Williams97,BarWil97,GolWilBis98,Sollich99,MalOpp01}. The
advantages of Gaussian processes are that prior assumptions about the
problem to be learned are encoded in a very transparent way, and that
inference---at least in the case of regression that we will
consider---is relatively straightforward; they are also
`non-parametric' in the sense that their effective number of
parameters (`degrees of freedom') can grow arbitrarily large as more
and more training data is collected.

One crucial question for applications is then how `fast' Gaussian
processes learn, \ie, how many training examples are needed to achieve
a certain level of generalization performance. The typical (as opposed
to worst case) behaviour is captured in the {\em learning curve},
which gives the average generalization error $\eps$ as a function of
the number of training examples $n$.  Several workers have derived
bounds on
$\eps(n)$~\cite{MicWah81,Plaskota90,Opper97,TreWilOpp99,OppViv99,WilViv00}
or studied its large $n$ asymptotics~\cite{Silverman85,Ritter96}. As
we will illustrate below, however, the existing bounds are often far
from tight; and asymptotic results will not necessarily apply for
realistic sample sizes $n$. Our main aim in this paper is therefore to
derive approximations to $\eps(n)$ which get closer to the true
learning curves than existing bounds, and apply both for small and
large $n$. We compare these approximations with existing bounds and
the results of numerical simulations; possible improvements to the
approximations are also discussed. Finally, we study an analytically
solvable example scenario which sheds light on how tight bounds on
learning curves can be made in principle. Summaries of the early
stages of this work have appeared in the conference
proceedings~\cite{Sollich99,Sollich99_ICANN_GP}.

In its simplest form, the regression problem that we are considering
is this: We are trying to learn a function $\teach$ which maps inputs
$x$ (real-valued vectors) to (real-valued scalar) outputs $\teach(x)$.
We are given a set of training data $D$, consisting of $n$
input-output pairs $(\xl,\yl)$; the training outputs $\yl$ may differ
from the `clean' target outputs $\teach(\xl)$ due to corruption by
noise. Given a test input $x$, we are then asked to come up with a
prediction $\stud(x)$ for the corresponding output, expressed either
in the simple form of a mean prediction $\meanstud(x)$ plus error
bars, or more comprehensively in terms of a `predictive distribution'
$P(\stud(x)|x,D)$. In a Bayesian setting, we do this by specifying a
prior $P(\stud)$ over our hypothesis functions, and a likelihood
$P(D|\stud)$ with which each $\stud$ could have generated the training
data; from this we deduce the posterior distribution
$P(\stud|D)\propto P(D|\stud)P(\stud)$.
%
\iffalse
 In the case of feedforward
networks, where the hypothesis functions $\stud$ are parameterized by
a set of network weights, the predictive distribution then needs to be
extracted by integration over this posterior, either by
computationally intensive Monte Carlo techniques or by approximations
which lead to analytically tractable integrals.
\fi
%
If we wanted to use a feedforward network for this task, we could
proceed as follows: Specify candidate networks by a set of weights
$w$, with prior probability $P(w)$. Each network defines a
(stochastic) input-output relation described by the distribution of
output $y$ given input $x$ (and weights $w$), $P(y|x,w)$. Multiplying
over the whole data set, we get the probability of the observed data
having been produced by the network with weights $w$:
$P(D|w)=\prod_{l=1}^n P(\yl|\xl,w)$. Bayes' theorem then gives us the
posterior, i.e., the probability of network $w$ given the data, as
$P(w|D) \propto P(D|w) P(w)$ up to an overall normalization
factor. From this, finally, we get the predictive distribution
$P(y|x,D)=\int dw P(y|x,w)P(w|D)$. This solves the regression problem
in principle, but leaves us with a nasty integral over all possible
network weights: the posterior $P(w|D)$ generally has a highly
nontrivial structure, with many local peaks (corresponding to local
minima in the training error). One therefore has to use sophisticated
Monte Carlo integration techniques~\cite{Neal93} or local
approximations to $P(w|D)$ around its maxima~\cite{MacKay92c} to
tackle this problem. Even once this has been done, one is still left
with the question of how to interpret the results: We may for example
want to select priors on the basis of the data, \eg\ by making the
prior $P(w|h)$ dependent on a set of hyperparameters $h$ and choosing
$h$ such as to maximize the probability $P(D|h)$ of the data. Once we
have found the `optimal' prior we would then hope that it tells us
something about the regression problem at hand (whether certain input
components are irrelevant, for example).  This would be easy if the
prior told us directly how likely certain input-output functions are;
instead we have to extract this information from the prior over
weights, often a complicated process.

\begin{figure}
\begin{center}
\epsfig{file=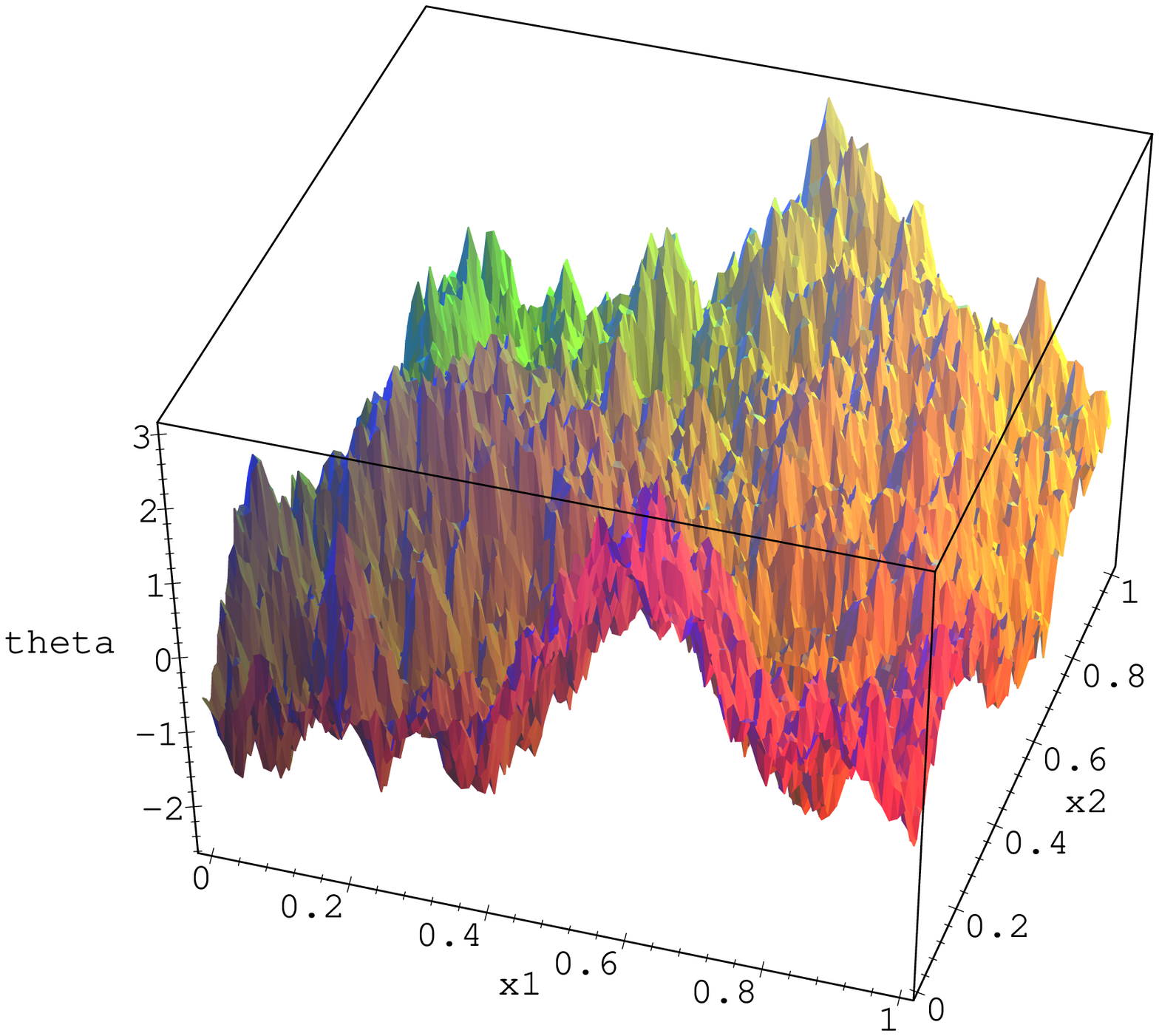, width=7cm}%
\hspace*{0.5cm}%
\epsfig{file=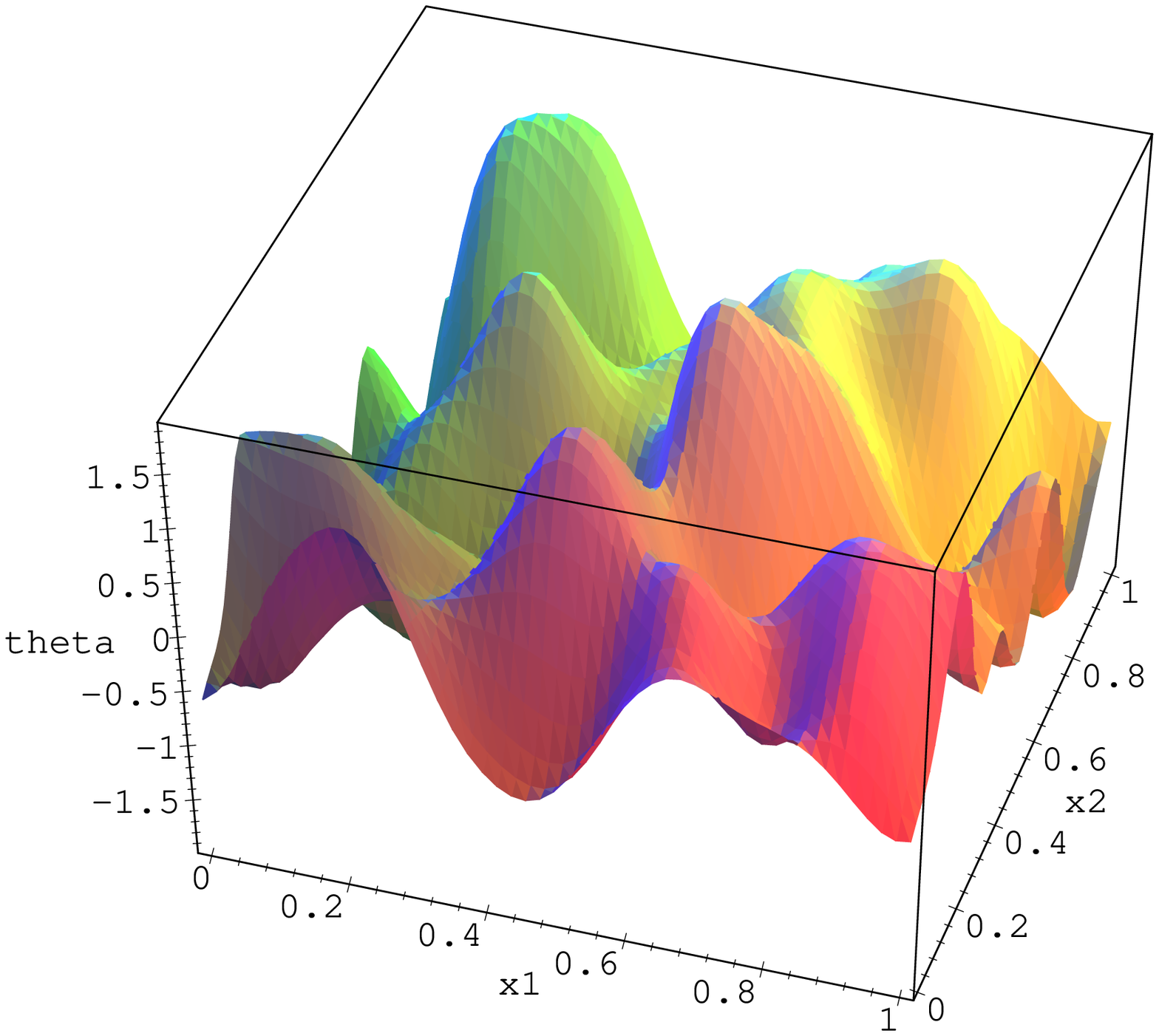, width=7cm}
\end{center}
\caption{Samples drawn from Gaussian process priors over functions on
$[0,1]^2$. Left: OU covariance function, $C(x,x')
=\exp(-||x-x'||/l)$. Right: RBF covariance function,
$C(x,x')=\exp[-||x-x'||^2/(2l^2)]$. The length scale $l=0.1$
determines in both cases over what distance the functions vary
significantly. Note the difference in roughness of the two functions;
this is related to the behaviour of the covariance functions for $x\to
x'$.
\label{fig:prior_samples}
}
\end{figure}

By contrast, for a Gaussian process it is an almost trivial task to
obtain the posterior and the predictive distribution (see below). One
reason for this is that the prior $P(\stud)$ is defined directly over
input-output functions $\stud$. How is this done? Any $\stud$ is
uniquely determined by its output values $\stud(x)$ for all $x$ from
the input domain, and for a Gaussian process, these are simply assumed
to have a joint Gaussian distribution (hence the name). This
distribution can be specified by the mean values
$\lav\stud(x)\rav_\stud$ (which we assume to be zero in the following,
as is commonly done), and the covariances
$\lav\stud(x)\stud(x')\rav_\stud = C(x,x')$; $C(x,x')$ is called the
{\em covariance function} of the Gaussian process. It encodes in an
easily interpretable way prior assumptions about the function to be
learned.  Smoothness, for example, is controlled by the behaviour of
$C(x,x')$ for $x'\to x$: The Ornstein-Uhlenbeck (OU) covariance
function $C(x,x') \propto \exp(-||x-x'||/l)$ produces very rough
(non-differentiable) functions, while functions sampled from the
radial basis function (RBF) prior with $C(x,x') \propto
\exp[-||x-x'||^2/(2l^2)]$ are infinitely differentiable. (Intermediate
priors yielding $r$ times differentiable functions can also be defined
by using modified Bessel functions as covariance functions;
see~\cite{WilViv00}.)  Figure~\ref{fig:prior_samples} illustrates
these characteristics with two samples from the OU and RBF priors,
respectively, over a two-dimensional input domain.  The `length scale'
parameter $l$ in the covariance functions also has an intuitive
meaning: It corresponds directly to the distance in input space over
which we expect our function to vary significantly. More complex
properties can also be encoded; by replacing $l$ with different length
scales for each input component, for example, relevant (small $l$) and
irrelevant (large $l$) inputs can be distinguished.

How does inference with Gaussian processes work? We only give a brief
summary here and refer to existing reviews on the subject (see
\eg~\cite{Williams98}) for details. It is simplest to assume that
outputs $y$ are generated from the `clean' values of a hypothesis
function $\stud(x)$ by adding Gaussian noise of $x$-independent
variance $\noise$. The joint distribution of a set of $n$ training outputs
$\{\yl\}$ and the function values $\stud(x)$ is then also Gaussian,
with covariances given by
\beastar
\lav \yl \ym \rav & = & C(\xl,\xm)+\noise\delta_{lm} = (\KK)_{lm}\\
\lav \yl \stud(x) \rav & = & C(\xl,x) \ \ = \ \ (\kk(x))_l
\eeastar
where we have defined an $n \times n$ matrix $\KK$ and an $x$-dependent
$n$-component vector $\kk(x)$. The posterior distribution
$P(\stud|D)$ is then obtained by simply conditioning on the $\{\yl\}$.
It is again Gaussian and has mean
\be
\hat\stud(x,D) \equiv \lav \stud(x) \rav_{\stud|D} = \kk(x)\T\KK^{-1}\yy
\label{basic_inf_a}
\ee
and variance
\be
\eps(x,D) \equiv \lav (\stud(x)-\hat\stud(x))^2 \rav_{\stud|D} =
C(x,x) - \kk(x)\T\KK^{-1}\kk(x).
\label{basic_inf_b}
\ee
Eqs.~(\ref{basic_inf_a},\ref{basic_inf_b}) solve the inference problem
for Gaussian process: They provide us directly with the predictive
distribution $P(\stud(x)|x,D)$. The posterior variance,
eq.~(\ref{basic_inf_b}), in fact also gives us the expected
generalization error (or Bayes error) at $x$. Why? If the teacher is
$\teach$, the squared deviation between our mean prediction and the
teacher output%
\footnote{%
One can also measure the generalization by the squared deviation
between the prediction $\hat\stud(x)$ and the {\em noisy} teacher
output; this simply adds a term $\noise$
to eq.~(\protect\ref{epsD_basic}).
}%
\ is $(\hat\stud(x)-\teach(x))^2$; averaging this over the posterior
distribution of teachers $P(\teach|D)$ just
gives~(\ref{basic_inf_b}). The underlying assumption is that our
assumed Gaussian process prior is the true one from which teachers are
actually generated (and that we are using the correct noise model).
Otherwise, the expected generalization error is larger and given by a
more complicated expression~\cite{WilViv00}. In line with most other
work on the subject, we only consider the `correct prior' case in the
following. Averaging the generalization error at $x$ over the
distribution of inputs gives then
\be
\eps(D) = \lav \eps(x,D) \rav_x
= \lav C(x,x) - \kk(x)\T\KK^{-1}\kk(x) \rav_x
\label{epsD_basic}
\ee
This form of the generalization error, which is well
known~\cite{MicWah81,Opper97,Williams98,WilViv00}, still depends on
the training inputs; the fact that the training {\em outputs} have dropped
out already is a signature of the fact that Gaussian processes are
{\em linear} predictors (compare~\(basic_inf_a)). Averaging over data
sets yields the quantity we are after,
\be
\eps = \lav\eps(D)\rav_D.
\label{eps_basic}
\ee
This average expected generalization error (we will drop the `average
expected' in the following) only depends on the number of training
examples $n$; the function $\eps(n)$ is called the {\em learning
curve}. Its exact calculation is difficult because of the joint
average in eqs.~(\ref{epsD_basic},\ref{eps_basic}) over the training
inputs $\xl$ and the test input $x$.

\begin{figure}
%\hspace*{-0.8cm}
\epsfig{file=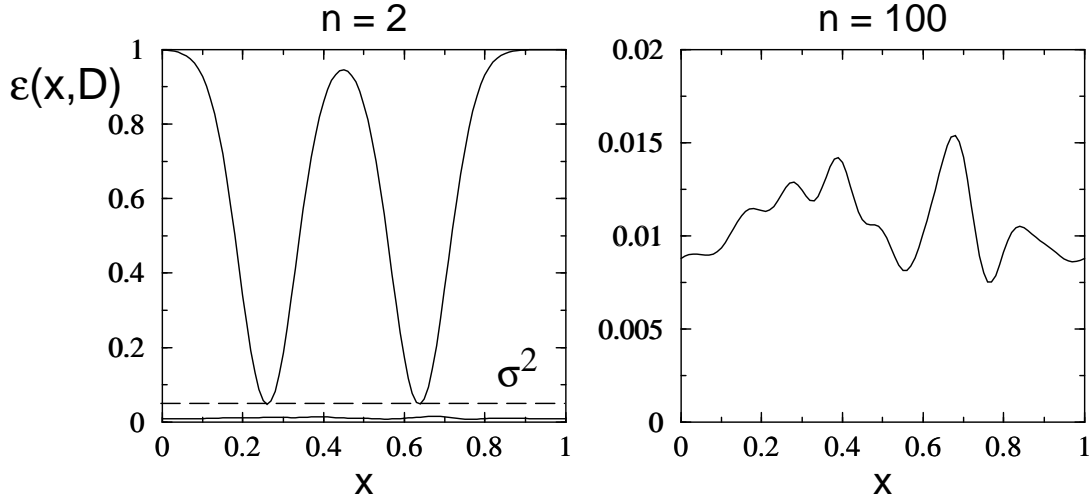, width=0.95\textwidth}
\caption{Generalization error $\eps(x,D)$ as a function of input
position $x\in[0,1]$, for noise level $\noise=0.05$, RBF covariance
function $C(x,x') = \exp[-|x-x'|^2/(2l^2)]$ with $l=0.1$, for randomly
drawn training sets $D$ of size $n=2$ (left) and $n=100$ (right).  To
emphasize the difference in scale, the plot on the left actually also
includes the results for $n=100$, just visible below the dashed line
at $\eps(x,D)=\noise$.
\label{fig:small_large_n}
}
\end{figure}

Before proceeding with our calculation of the learning curve
$\eps(n)$, let us try to gain some intuitive insight into its
dependence on $n$. Consider a simple example scenario, where inputs
$x$ are one-dimensional and drawn randomly from the unit interval
$[0,1]$, with uniform probability. For the covariance function we
choose an RBF form, $C(x,x') = \exp[-|x-x'|^2/(2l^2)]$ with
$l=0.1$. Here we have taken the prior variance $C(x,x)$ as unity; as
seems realistic for most applications we assume the noise level to be
much smaller than this, $\noise =
0.05$. Figure~\ref{fig:small_large_n} illustrates the $x$-dependence
of the generalization error $\eps(x,D)$ for a small training set
($n=2$): Each of the examples has made a `dent' in $\eps(x,D)$, with a
shape that is similar to that of the covariance function%
\footnote{%
More precisely, the dents have the shape of the {\em square} of the
covariance function: If the training inputs $x_i$ are sufficiently far
apart, then around each $x_i$ we can neglect the influence of the
other data points and apply~(\protect\ref{basic_inf_b}) with $n=1$,
giving $\eps(x,D)\approx C(x,x) - C^2(x,x_i)/[C(x_i,x_i)+\noise]$.
}%
. Outside the dents, $\eps(x,D)$ still has essentially its prior
value, $\eps(x,D)=1$; at the centre of each dent it is reduced to a
much smaller value, $\eps(x,D)\approx \noise/(1+\noise)$ (this
approximation holds as long as the different training inputs are
sufficiently far away from each other). The generalization error
$\eps(D)$ is therefore dominated by regions where no training examples
have been seen; one has $\eps(D)\gg \noise$, and the precise value of
$\eps(D)$ depends only very little on $\noise$ (assuming always that
$\noise\ll 1$).  Gradually, as $n$ is increased, the covariance dents
will cover the input space, so that $\eps(x,D)$ and $\eps(D)$ become
of order $\noise$; this situation is shown on the right of
Figure~\ref{fig:small_large_n}. From this point onwards, further
training examples essetially have the effect of averaging out noise,
eventually making $\eps(D)\ll \noise$ for large enough $n$. In
summary, we expect the learning curve $\eps(n)$ to have two regimes:
In the initial (small $n$) regime, where $\eps(n)\gg \noise$,
$\eps(n)$ is essentially independent of $\noise$ and reflects mainly
the geometrical distribution of covariance dents across the inputs
space. In the asymptotic regime ($n$ large enough such that
$\eps(n)\ll\noise$), on the other hand, the noise level $\noise$ is
important in controlling the size of $\eps(n)$ because learning arises
mainly from the averaging out of noise in the training data.

%

\iffalse
Review of previous results: asymptotics,
Plaskota (useful mainly in small noise limit), Chris' upper bound (for
first part of learning curve), Manfred's bounds (second part of
learning curve).
\fi

\section{Approximate learning curves}

\iffalse
Starting point: Eigenvalue decomposition of covariance function, and
then Woodbury.
\fi

Calculating learning curves for Gaussian processes exactly is a
difficult problem because of the joint average
in~(\ref{epsD_basic},\ref{eps_basic}) over the training inputs $\xl$
and the test input $x$. Several workers have therefore derived upper
and lower bounds on $\eps$~\cite{MicWah81,Plaskota90,Opper97,WilViv00}
or studied the large $n$ asymptotics of
$\eps(n)$~\cite{Silverman85,Ritter96}. As we will illustrate below,
however, the existing bounds are often far from tight; likewise,
asymptotic results can only capture the large $n$ regime defined above
and will not necessarily apply for sample sizes $n$ occurring in
practice. We therefore now attempt to derive approximations to
$\eps(n)$ which get closer to the true learning curves than existing
bounds, and which are applicable both for small and large $n$.

As a starting point for an approximate calculation of $\eps(n)$, we
first derive a representation of the generalization error in terms of
the eigenvalue decomposition of the covariance function. Mercer's
theorem (see \eg~\cite{Wong71}) tells us that the covariance function
can be decomposed into its eigenvalues $\lam_i$ and eigenfunctions
$\phi_i(x)$:
\be
C(x,x')= \sum_{i=1}^\infty \lam_i \phi_i(x) \phi_i(x')
\label{c_decomp}
\ee
This is simply the analogue of the eigenvalue decomposition of a
finite symmetric matrix. We assume here that eigenvalues and
eigenfunctions are defined relative to the distribution over inputs
$x$, \ie,
\be
\lav C(x,x')\phi_i(x')\rav_{x'} = \lam_i \phi_i(x)
\label{eigenfunc}
\ee
The eigenfunctions are then orthogonal with respect to the same
distribution, $\lav \phi_i(x)\phi_j(x)
\rav_x=\delta_{ij}$ (see \eg~\cite{WilSee00}). Now
write the data-dependent generalization error~\(epsD_basic) as
\[
\eps(D) = \lav C(x,x) \rav_x -
\tr \lav\kk(x)\kk(x)\T\rav_x \KK^{-1}
\]
and perform the $x$-average:
\[
\lav (\kk(x)\kk(x)\T)_{lm} \rav_x = 
\sum_{ij} \lam_i \lam_j 
\phi_i(\xl) \lav\phi_i(x)\phi_j(x)\rav_x \phi_j(\xm)
= \sum_i \lam_i^2 \phi_i(\xl) \phi_i(\xm).
\]
This suggests introducing the diagonal matrix
$(\LL)_{ij}=\lam_i \delta_{ij}$ and the `design
matrix' $(\Ph)_{li}=\phi_i(\xl)$, so that $\lav
\kk(x)\kk(x)\T \rav_x = \Ph\LL^2\Ph\T$. One then also has
\be
\lav C(x,x) \rav_x = \tr \LL
\label{Cav}
\ee
and the matrix $\KK$ is expressed as
\[
\KK = \noise\mident+\Ph\LL\Ph\T
\]
with $\mident$ being the identity matrix. Collecting these results, we
have
\[
\eps(D) = \tr\LL - \tr (\noise\mident+\Ph\LL\Ph\T)^{-1}\Ph\LL^2\Ph\T
\]
This can be simplified using the Woodbury formula%
\footnote{%
$(\ma + \mU\mv\T)^{-1} = \ma^{-1} - \ma^{-1}\mU(\mident +
\mv\T\ma^{-1}\mU)^{-1} \mv\T\ma^{-1}$ for matrices $\ma$, $\mU$, $\mv$
of appropriate size; see \eg~\protect\cite{PreTeuVetFla92}.
}
to give%
\footnote{%
If the covariance function has zero eigenvalues, the inverse
$\LL^{-1}$ does not exist, and the first form of $\eps(D)$ given
in~(\protect\ref{basic_result}) must be used; similar alternative
forms, though not explicitly written, exist for all following
results.}
$\eps(n) = \lav\eps(D)\rav_D$ with
\be
\eps(D) = 
\tr \LL(\mident+\noiseinv\LL\Ph\T\Ph)^{-1} = 
\tr (\LL^{-1}+\noiseinv\Ph\T\Ph)^{-1}
\label{basic_result}
\ee
The advantage of this (still exact) representation of the
generalization error is that the average over the test input $x$ has
already been carried out, and that the remaining dependence on the
training data is contained entirely in the matrix $\Ph\T\Ph$.  It also
includes as a special case the well-known result for linear regression
(see \eg~\cite{Sollich94f}); $\LL^{-1}$ and $\Ph\T\Ph$ can be
interpreted as suitably generalized versions of the weight decay
(matrix) and input correlation matrix.

Starting from~\(basic_result), one can now derive approximate
expressions for the learning curve $\eps(n)$.  The most naive approach
is to neglect entirely the fluctuations in $\Ph\T\Ph$ over different
data sets and replace it by its average, which is simply $\lav
(\Ph\T\Ph)_{ij} \rav_D = \sum_{l=1}^n \lav
\phi_i(\xl)\phi_j(\xl)\rav_D = n \delta_{ij}$. This leads to
\be
\epsOV(n) = \tr (\LL^{-1} + \noiseinv n\mident)^{-1}.
\label{epsN}
\ee
While this is not, in general, a good {\em approximation}, it was
shown by {\em O}pper and {\em V}ivarelli to be a lower
{\em bound} (called OV bound below) on the learning
curve~\cite{OppViv99}. It becomes tight in the large noise limit
$\noise\to\infty$ at constant $n/\noise$: The fluctuations of the
elements of the matrix $\noiseinv\Ph\T\Ph$ then become vanishingly
small (of order $\sqrt{n}\noiseinv=(n/\noise)/\sqrt{n}\to 0$) and so
replacing $\Ph\T\Ph$ by its average is justified.

To derive better approximations, it is useful to see how the matrix
$\Gfluc=(\LL^{-1}+\noiseinv\Ph\T\Ph)^{-1}$ changes when a new example
is added to the training set. This change can be expressed as
\bea
\Gfluc(n\!+\!1)-\Gfluc(n)&=&
\left[\Gfluc^{-1}(n)+
\noiseinv\new\new\T\right]^{-1}-\Gfluc(n)
\nonumber\\
&=&
-\,\frac
{\Gfluc(n) \new\new\T\Gfluc(n)} {\noise+\new\T\Gfluc(n)\new}
\label{G_update}
\eea
in terms of the vector $\new$ with elements $(\new)_i =
\phi_i(x_{n+1})$. To get exact learning curves, one would have to
average this update formula over both the new training input $x_{n+1}$
and all previous ones. This is difficult, but progress can be made by
neglecting correlations of numerator and denominator in~\(G_update),
averaging them separately instead. Also treating $n$ as a continuous
variable, this yields the approximation
\be
\frac{\partial\GG(n)}{\partial n} = 
-\, \frac{\lav\Gfluc^2(n)\rav}{\noise+\tr\GG(n)}
\label{avG_update}
\ee
where we have introduced the notation $\GG=\lav\Gfluc\rav$. If we
also neglect fluctuations in $\Gfluc$, approximating
$\lav\Gfluc^2\rav=\GG^2$, this equation can easily be solved and yields
$\GG^{-1}(n)=\LL^{-1}+\noiseinv \nUC \mident$ and so
\be
\epsUC(n) = \tr (\LL^{-1} +\noiseinv\nUC\mident)^{-1}
\label{epsUC}
\ee
with $\nUC$ determined by the self-consistency equation
\[
\nUC + \tr
\ln (\mident+\noiseinv \nUC \LL) = n.
\]
By comparison with~\(epsN), $\nUC$ can be thought of as an `effective
number of training examples'. The subscript UC in~\(epsUC) stands for
{\em U}pper {\em C}ontinuous (\ie, treating $n$ as continuous)
approximation. A better approximation with a lower value is obtained
by retaining fluctuations in $\Gfluc$. As in the case of the linear
perceptron~\cite{Sollich94f}, this can be achieved by introducing an
auxiliary offset parameter $v$ into the definition of
\be
\Gfluc^{-1}=v\mident+\LL^{-1}+\noiseinv\Ph\T\Ph.
\label{vdef}
\ee
One can then write
\be
-\tr\lav\Gfluc^2\rav=\frac{\partial}{\partial v}\tr\lav\Gfluc\rav =
\partial\eps/\partial v
\label{Gsquared}
\ee
and obtains from\eq{avG_update} the partial differential
equation
\be
\frac{\partial\eps}{\partial n}
- \frac{1}{\noise+\eps}\frac{\partial\eps}{\partial v}
= 0
\label{LC_eqn}
\ee
This can be solved for $\eps(n,v)$ using the methods of characteristic
curves (see App.~\ref{app:char}). Resetting the auxiliary parameter
$v$ to zero yields the {\em L}ower {\em C}ontinuous approximation to
the learning curve, which is given by the self-consistency equation
\be
\epsLC(n) =
\tr\left(\LL^{-1}+\frac{n}{\noise+\epsLC}\mident\right)^{-1}.
\label{epsLC}
\ee
It is easy to show that $\epsLC\leq\epsUC$. One can also check that
both approximations converge to the exact result~\(epsN) in the large
noise limit (as defined above). Encouragingly, we see that the LC
approximation reflects our intuition about the difference between the
initial and asymptotic regimes of the learning curve: For
$\eps\gg\noise$, we can simplify\eq{epsLC} to
\[
\epsLC(n) =
\tr\left(\LL^{-1}+\frac{n}{\epsLC}\mident\right)^{-1}
\]
where as expected the noise level $\noise$ has dropped out. In the
opposite limit $\eps\ll\noise$, on the other hand, we have
\be
\epsLC(n) = 
\tr\left(\LL^{-1}+\frac{n}{\noise}\mident\right)^{-1}
\label{epsLC_nlarge}
\ee
which---again as expected---retains the noise level $\noise$ as an
important parameter. Eq.\eq{epsLC_nlarge} also shows that $\epsLC$
approaches the OV lower bound (from above) for sufficiently large $n$.

We conclude this section with a brief qualitative discussion of the
expected $n$-dependence of $\epsLC$ (which below will turn out to be
the more accurate of our two approximations). Obviously this
$n$-dependence depends on the spectrum of eigenvalues $\lambda_i$;
below we always assume that these are arranged in decreasing order.
Consider then first the asymptotic regime $\eps\ll\noise$, where
$\epsLC$ and $\epsOV$ become identical. One then shows easily that for
eigenvalues decaying as a power-law, $\lam_i \sim i^{-r}$, the
asymptotic learning curve scales as%
\footnote{
Among the scenarios studied in the next section, the OU covariance
function provides a concrete example of this kind of behaviour: In
$d=1$ dimension it has, from~(\protect\ref{OU_eigenvals}), $\lam_i\sim
i^{-2}$ and thus $r=2$. The RBF covariance function, on the other
hand, has eigenvalues decaying faster than any power law,
corresponding to $r\to\infty$ and thus $\epsLC\sim \noise/n$ (up to
logarithmic corrections) in the asymptotic regime.
}
 $\epsLC\sim (n/\noise)^{-(r-1)/r}$; this is in agreement with known
exact results~\cite{Silverman85,Ritter96}. In the initial regime
$\eps\gg\noise$, on the other hand, one can take $\noise\to 0$ and
finds then a faster decay%
\footnote{
For covariance functions with eigenvalues decaying faster than a power
law, the behaviour in the initial regime is nontrivial; for the RBF
covariance function in $d=1$ with uniform inputs, for example, we find
(for large $n$) $\epsLC \sim n\exp(-cn^2)$ with some constant $c$.
}
 of the generalization error, $\epsLC\sim n^{-(r-1)}$. We are not
aware of exact results pertaining to this regime, except for the OU
case in $d=1$ which has $r=2$ and for which thus $\epsLC \sim n^{-1}$,
in agreement with an exact calculation (Manfred Opper, private
communication).

\section{Comparisons with bounds and numerical simulations}
\label{sec:comparison}

We now compare the LC and UC approximations with existing bounds, and
with the `true' learning curves as obtained by numerical
simulations. A lower bound on the generalization error was given by
Michelli and Wahba~\cite{MicWah81} as
\be
\eps(n)\geq\epsMW(n)=\sum_{i=n+1}^{\infty} \lam_i
\label{epsMW}
\ee
This bound is derived for the noiseless case by allowing `generalized
observations' (projections of $\teach(x)$ along the first $n$
eigenfunctions of $C(x,x')$), and so is unlikely to be tight for the
case of `real' observations at discrete input points. Given that the
bound is derived from the $\noise\to 0$ limit, it can only be useful
in the initial (small $n$, $\eps\gg\noise$) regime of the learning
curve. There, it confirms the conclusion of our intuitive discussion
above that the learning curve has a lower limit below which it will
not drop even for $\noise\to 0$.

Plaskota~\cite{Plaskota90} generalized the MW approach to the noisy
case and obtained the following improved lower bound:
\be
\eps(n)\geq\epsPl(n) = \min_{\{\eta_i\}} \sum_{i=1}^n
\frac{\lam_i\noise}{\eta_i+\noise} + \sum_{i=n+1}^{\infty} \lam_i
\label{Plaskota}
\ee
where the minimum is over all non-negative $\eta_1, \ldots, \eta_n$
obeying $\sum_{i=1}^n\eta_i$ $=$ $S$ $\equiv$ $\sum_{l=1}^n
C(\xl,\xl)$. Plaskota only derived this bound for covariance functions
for which the prior variance $C(x,x)$ is independent of $x$. We call
these `uniform' covariance functions; due to the general identity
$\lav C(x,x)\rav_x = \tr\LL$, they obey $C(x,x)=\tr\LL$ (and hence
$S=n\,\tr\LL$). We recap Plaskota's proof in App.~\ref{app:Plaskota}
and also show there that it extends to general covariance functions in
the form stated above. The Plaskota bound is close to the MW bound in
the small $n$ regime (equivalent to $\noise\to0$); for larger $n$ it
becomes substantially larger. It therefore has the potential to be
useful for both small and large $n$. Note that, in contrast to all
other bounds discussed in this paper, the MW and Plaskota bounds are
in fact `single data set' (worst case) bounds: They apply to $\eps(D)$
for {\em any} data set $D$ of the given size $n$, rather than just to
the average%
\footnote{%
Our generalized version of the Plaskota bound depends on the specific
data set only through the value of $S=\sum_{l=1}^n C(\xl,\xl)$. To
obtain an average case bound one would need to average over the
distribution of $S$.
}%
\ $\eps(n)=\lav\eps(D)\rav$.

Opper used information theoretic methods to obtain a different lower
bound~\cite{Opper97}, but we will not consider this because the more
recent OV bound~\(epsN) is always tighter. Note that the OV bound
incorrectly suggest that $\eps$ decreases to zero for $\noise\to 0$ at
fixed $n$. It therefore becomes void for small $n$ (where
$\eps\gg\noise$) and is expected to be of use only in the asymptotic
regime of large $n$.

There is also an {\em U}pper bound due to {\em O}pper~\cite{Opper97},
\be
\tilde\eps(n)\leq \epsUO(n)=
(\noiseinv n)^{-1}\,\tr\ln(\mident + \noiseinv n \LL) + 
\tr (\LL^{-1}+\noiseinv n\mident)^{-1}
\label{epsUO}
\ee
Here $\tilde\eps$ is a modified version of $\eps$ which (in the
rescaled version that we are using) becomes identical to $\eps$ in the
limit of small generalization errors ($\eps\ll\noise$), but never gets
larger that $2\noise$; for small $n$ in particular, $\eps(n)$ can
therefore actually be much {\em larger} than $\tilde\eps(n)$ and its
bound~\(epsUO). For this reason, and because in our simulations we
never get very far into the asymptotic regime $\eps\ll\noise$, we do
not display the UO bound in the graphs below.

The UO bound is complemented by an upper bound due to {\em W}illiams
and {\em V}ivarelli~\cite{WilViv00}, which never decays below values
around $\noise$ and is therefore mainly useful in the initial regime
$\eps\gg\noise$. It applies for one-dimensional inputs $x$ and
stationary covariance functions---for which $C(x,x')=C_s(x-x')$ is a
function of $x-x'$ alone---and reads:
\be
\eps(n)\leq \epsWV(n) = C_s(0) - \frac{1}{C_s(0)+\noise}
\int_0^\infty\! da\,
f_n(a)\, C_s^2(a)
\label{epsWV}
\ee
with
\be
f_n(a) = 2(1-a)^n \Theta(1-a) + 2(n-1)(1-2a)^n \Theta(1-2a)
\label{epsWVunif}
\ee
and where the Heaviside step functions $\Theta$ (defined as
$\Theta(z)=1$ for $z>0$ and $=0$ otherwise) in the two terms imply
that only values of $a$ up to 1 and $1/2$, respectively, contribute to
the integral in\eq{epsWV}. The function $f_n(a)$ is a normalized
distribution over $a$ which for $n\to\infty$ becomes peaked around
$a=0$, implying that the asymptotic value of the bound is
$\epsWV(n\to\infty) = C_s(0)\noise/[C_s(0)+\noise]$ $\approx$ $\noise$
for $\noise\ll C_s(0)$. The derivation of the bound is based on the
insight that $\eps(x,D)$ always decreases as more examples are added;
it can therefore be upper bounded for any given $x$ by the smallest
$\eps(x,D')$ that would result from training on any data set $D'$
comprising only a {\em single} example from the original training set
$D$. The idea can be generalized to using the smallest $\eps(x,D')$
obtainable from any {\em two} of the training examples, but this does
not significantly improve the bound~\cite{WilViv00}.

As stated above in\eqq{epsWV}{epsWVunif}, the WV bound
applies only to the case of a uniform input distributions over the
unit interval $[0,1]$. However, it is relatively straightforward to
extend the approach to general (one-dimensional) input distributions
$P(x)$; only the data set average becomes technically a little more
complicated. We omit the details and only quote the result:
Eq.\eq{epsWV} remains valid if the expression\eq{epsWVunif} for
$f_n(a)$ is generalized to
\bea
f_n(a) &=& n \lav P(x-a)[1-Q(x)]^{n-1} + P(x+a) [Q(x)]^{n-1} \rav_x 
\nonumber\\
& & {}+{}
 n(n-1) \lav \Theta(x'-x-2a) [P(x+a)+P(x'-a)] [1+Q(x)-Q(x')]^{n-2} \rav_{x,x'}
\label{epsWVnonunif}
\eea
where $Q(z)=\int_{-\infty}^z \! dx\ P(x)$ is the cumulative
distribution function. In the simpler scenario considered by Williams
and Vivarelli this can be shown to reduce to\eq{epsWVunif}, while in
the most general case the numerical evaluation of the bound requires a
triple integral (over $x$, $x'$ and $a$).

Finally, there is one more upper bound, due to {\em T}recate, {\em
W}illiams and {\em O}pper~\cite{TreWilOpp99}; based on the
generalization error achieved by a `suboptimal' Gaussian regressor,
they showed
\[
\eps(n)\leq \epsTWO(n) = \tr\LL - n\sum_i \frac{\lam_i}{c_i}
\]
where
\be
c_i = (n-1)\lam_i + \noise + \lav C(x,x)\phi_i^2(x)\rav_x.
\label{epsTWO_ci}
\ee
For a uniform covariance function, the average in $c_i$ becomes
$\tr\LL\lav\phi_i^2(x)\rav_x=\tr\LL$ and the bound simplifies to
\[
\epsTWO(n) = \sum_i \lam_i\, \frac{\tr\LL + \noise - \lam_i}
{\tr\LL + \noise + (n-1)\lam_i}
\]

We now compare the quality of these bounds and our approximations with
numerical simulations of learning curves. All the theoretical
expressions require knowledge of the eigenvalue spectrum of the
covariance function, so we focus on situations where this is known
analytically. We consider three scenarios: For the first two, we
assume that inputs $x$ are drawn from the $d$-dimensional unit
hypercube $[0,1]^d$, and that the input density is uniform. As
covariance functions we use the RBF function
$C(x,x')=\exp[-||x-x'||^2/(2l^2)]$ and the OU function
$C(x,x')=\exp(-||x-x'||/l)$, to have the extreme cases of smooth and
rough functions to be learned; both contain a tunable length scale
$l$. To be precise, we use slightly modified versions of the RBF and
OU covariance functions (using what physicists call `periodic boundary
conditions') which make the eigenvalue calculations analytically
tractable; the details are explained in App.~\ref{app:spectra}. In the
third scenario we explore the effect of a non-uniform input
distribution by considering inputs $x$ drawn from a $d$-dimensional
(zero mean) isotropic Gaussian distribution $P(x)\propto
\exp[-||x||^2/(2\sigma_x^2)]$, for an RBF covariance function. Details
of the eigenvalue spectrum for this case can also be found in
App.~\ref{app:spectra}. Note that in all three cases, the covariance
function is uniform, \ie\ has a constant variance $C(x,x)$; we have
fixed this to unity without loss of generality. This leaves three
variable parameters: the input space dimension $d$, the noise level
$\noise$ and the length scale $l$. As explained above we generically
expect the prior variance to be significantly larger than the noise on
the training data, so we only consider values of $\noise<1$. The
length scale $l$ should also obey $l<1$; otherwise the covariance
functions $C(x,x')$ would be almost constant across the input space,
corresponding to a trivial GP prior of essentially $x$-independent
functions. We in fact choose the length scale $l$ for each $d$ in such
a way as to get a reasonable decay of the learning curve within the
range of $n=0\ldots 300$ that can be conveniently simulated
numerically. To see why this is necessary, note that each covariance
`dent' covers a fraction of order $l^d$ of the input space, so that
the number of examples $n$ needed to see a significant reduction in
generalization error $\eps$ will scale as $(1/l)^d$. This quickly
becomes very large as $d$ increases unless $l$ is increased
simultaneously. (The effect of larger $l$ leading to a faster decay of
the learning curve was also observed in~\cite{WilViv00}.)

In the following figures, we show the lower bounds (Plaskota, OV),
the non-asymptotic upper bounds (TWO and, for $d=1$ with uniform input
distribution, WV), and our approximations (LC and UC). The true
learning curve as obtained from numerical simulations is also
shown. For the numerical simulations, we built up training sets by
randomly drawing training inputs from the specified input
distribution. For each new training input, the matrix inverse
$\KK^{-1}$ has to be recalculated. By partitioning the matrix into its
elements corresponding to the old and new inputs, this inversion can
be performed with $\order(n^2)$ operations (see
\eg~\cite{PreTeuVetFla92}), as opposed to $\order(n^3)$ if the
inverse is calculated from scratch every time. With $\KK^{-1}$
known, the generalization error $\eps(D)$ was then calculated
from\eq{epsD_basic}, with the average over $x$ estimated by an average
over randomly sampled test inputs. This process was repeated up to our
chosen $n_{\rm max}=300$; the results for $\eps(D)$ were then averaged
over a number of training set realizations to obtain the learning
curve $\eps(n)$. In all the graphs shown, the size of the error bars
on the simulated learning curve is of the order of the visible
fluctuations in the curve.

\begin{figure}
\begin{center}
\epsfig{file=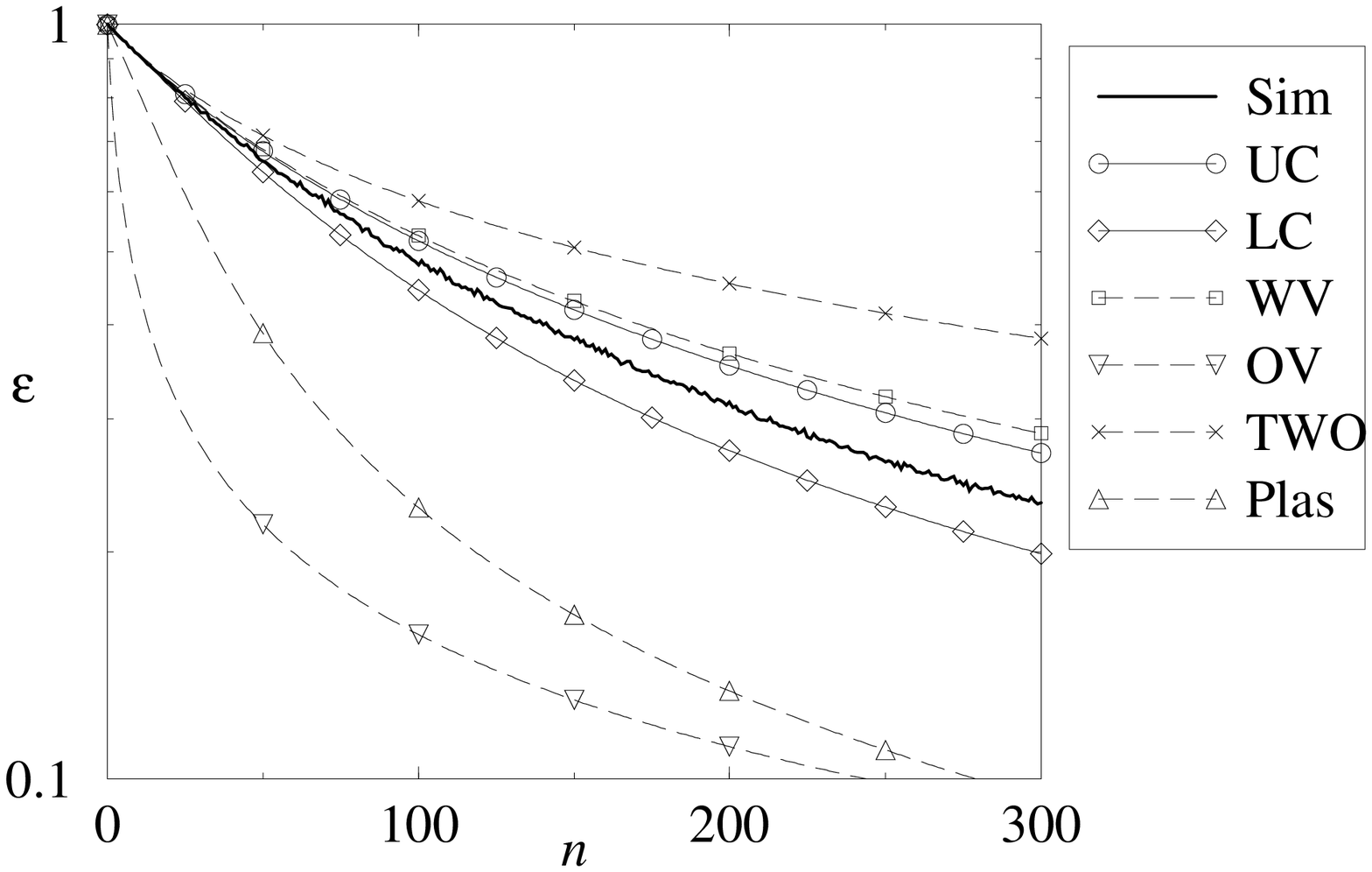, height=5.5cm}%
\hspace*{0.2cm}%
\epsfig{file=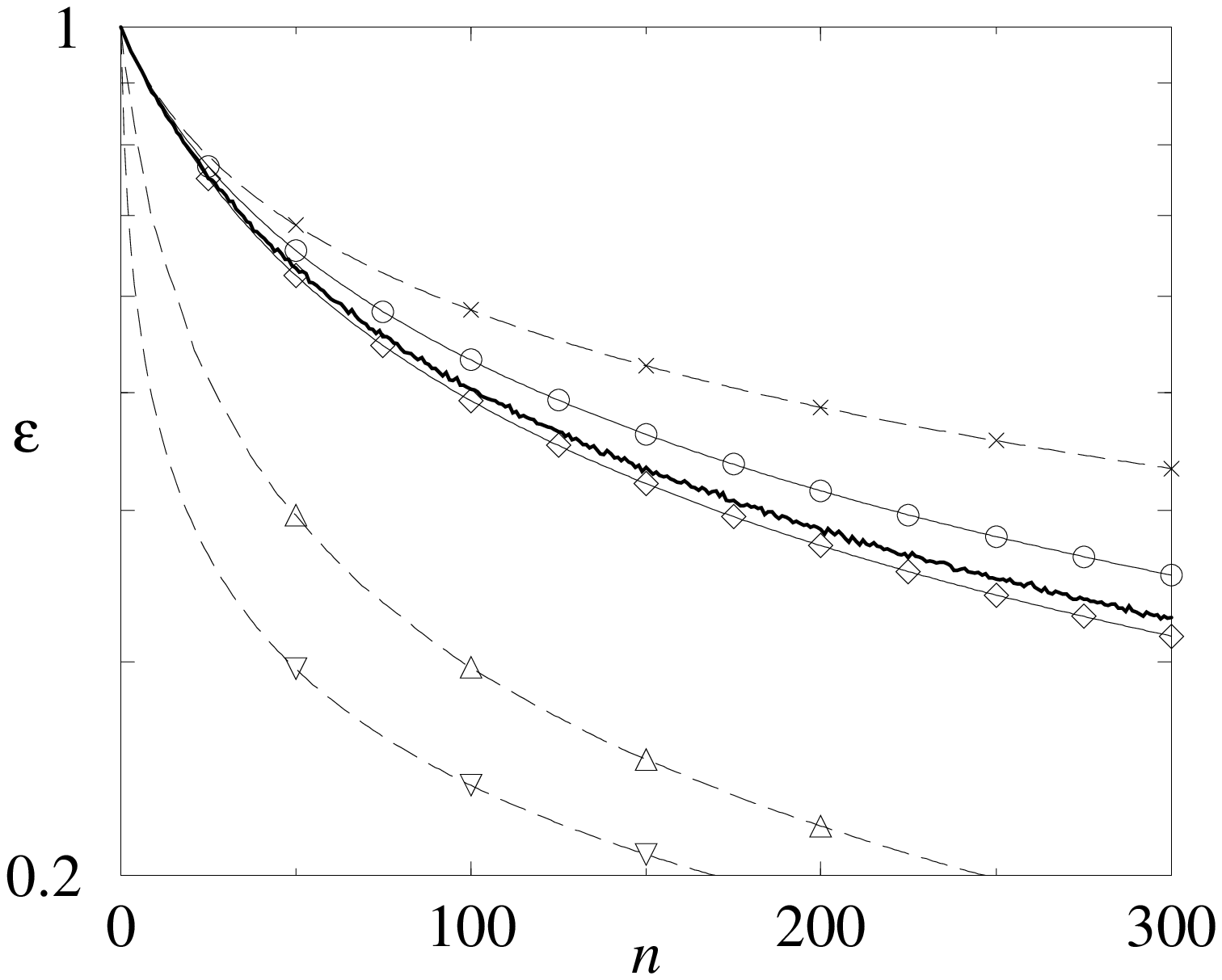, height=5.5cm}
\end{center}
\caption{Learning curve for a GP with OU covariance function and
inputs uniformly drawn from $x\in[0,1]^d$, at noise level
$\noise=0.05$. Left: $d=1$, length scale $l=0.01$. Right: $d=2$, $l=0.1$.
\label{fig:OUcube}
}
\end{figure}
In Figure~\ref{fig:OUcube} we show the results for an OU covariance
function with inputs from $[0,1]^d$, for $d=1$ (left) and $d=2$
(right). One observes that the lower bounds (Plaskota and OV) are
rather loose in both cases. The TWO upper bound is also far from
tight; the WV upper bound is better where it can be defined (for
$d=1$). Our approximations, LC and UC, are closer to the true learning
curve than any of the bounds and in fact appear to bracket it.

\begin{figure}
\begin{center}
\epsfig{file=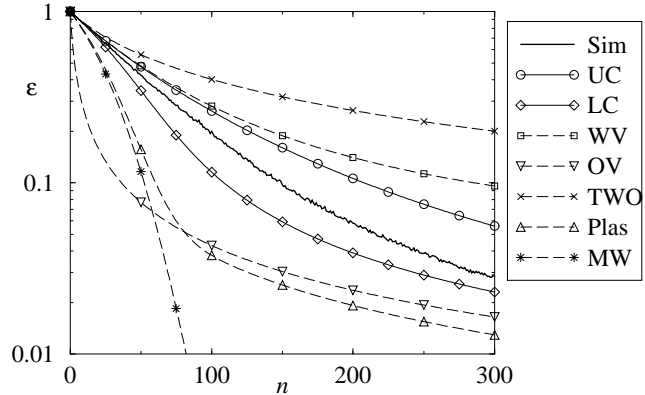, height=5.5cm}
\end{center}
\caption{Learning curve for a GP with RBF covariance function and
inputs uniformly drawn from $x\in[0,1]^d$, at noise level
$\noise=0.05$; dimension $d=1$, length scale $l=0.01$. 
We also show the MW bound here to show how it is first close to the
Plaskota bound but then ``misses'' the change in behaviour where the
generalization error crosses over into the asymptotic regime
($\eps\ll\noise$). 
\label{fig:RBFcube_oned}
}
\end{figure}
Similar comments apply to Figure~\ref{fig:RBFcube_oned}, which
displays the results for an RBF covariance function with inputs from
$[0,1]$. Because functions from an RBF prior are much smoother than
those from an OU prior, they are easier to learn and the
generalization error $\eps$ shows a more rapid decrease with $n$. This
makes visible, within the range of $n$ shown, the anticipated change
in behaviour as $\eps$ crosses over from the initial ($\eps\gg\noise$)
to the asymptotic ($\eps\ll\noise$) regime. The LC approximation and,
to a less quantitative extent, the Plaskota bound both capture this
change. By contrast, the OV bound (as expected from its general
properties discussed above) only shows the right qualitative behaviour
in the asymptotic regime.

\begin{figure}
\begin{center}
\epsfig{file=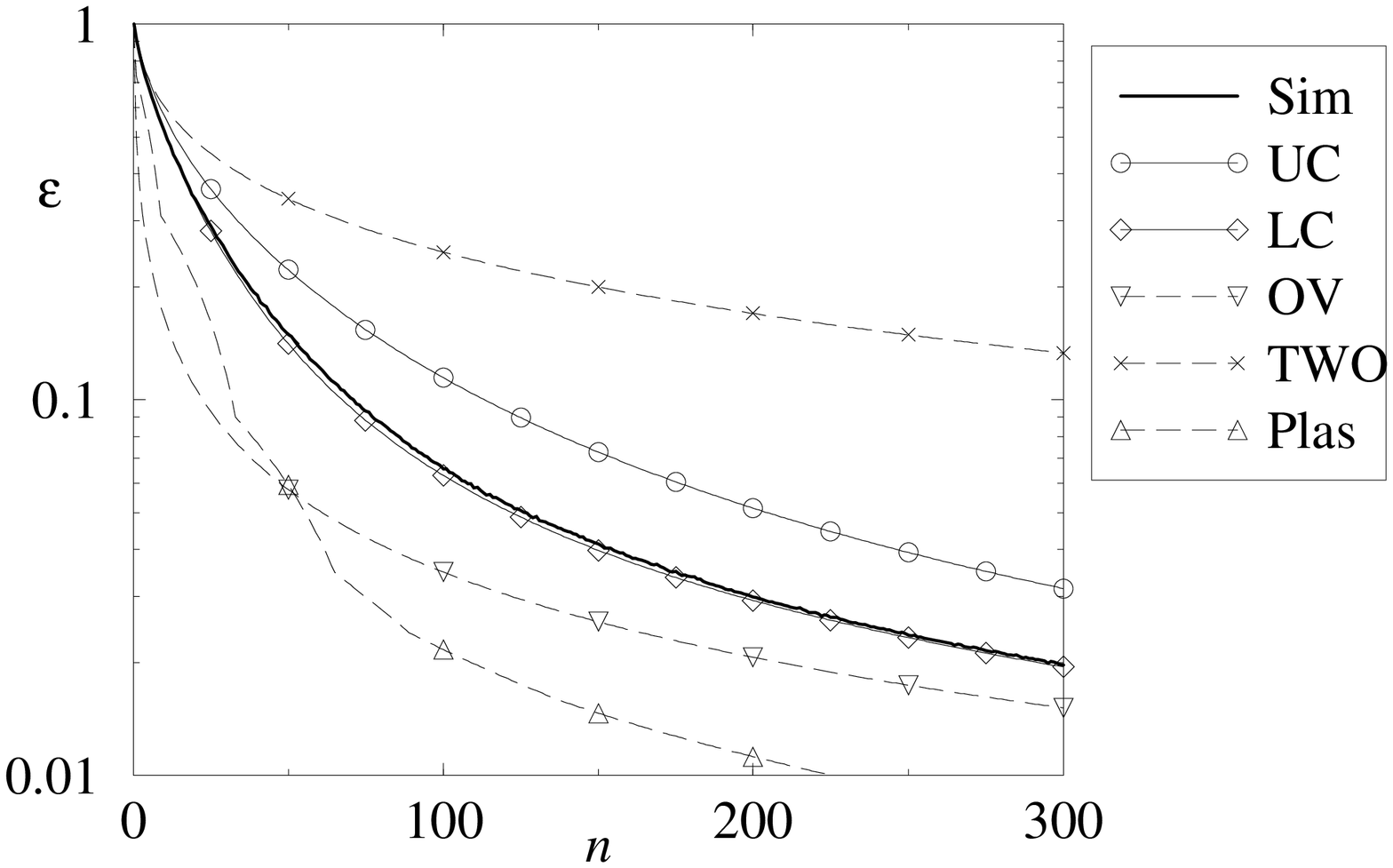, height=5.5cm}%
\hspace*{0.05cm}%
\epsfig{file=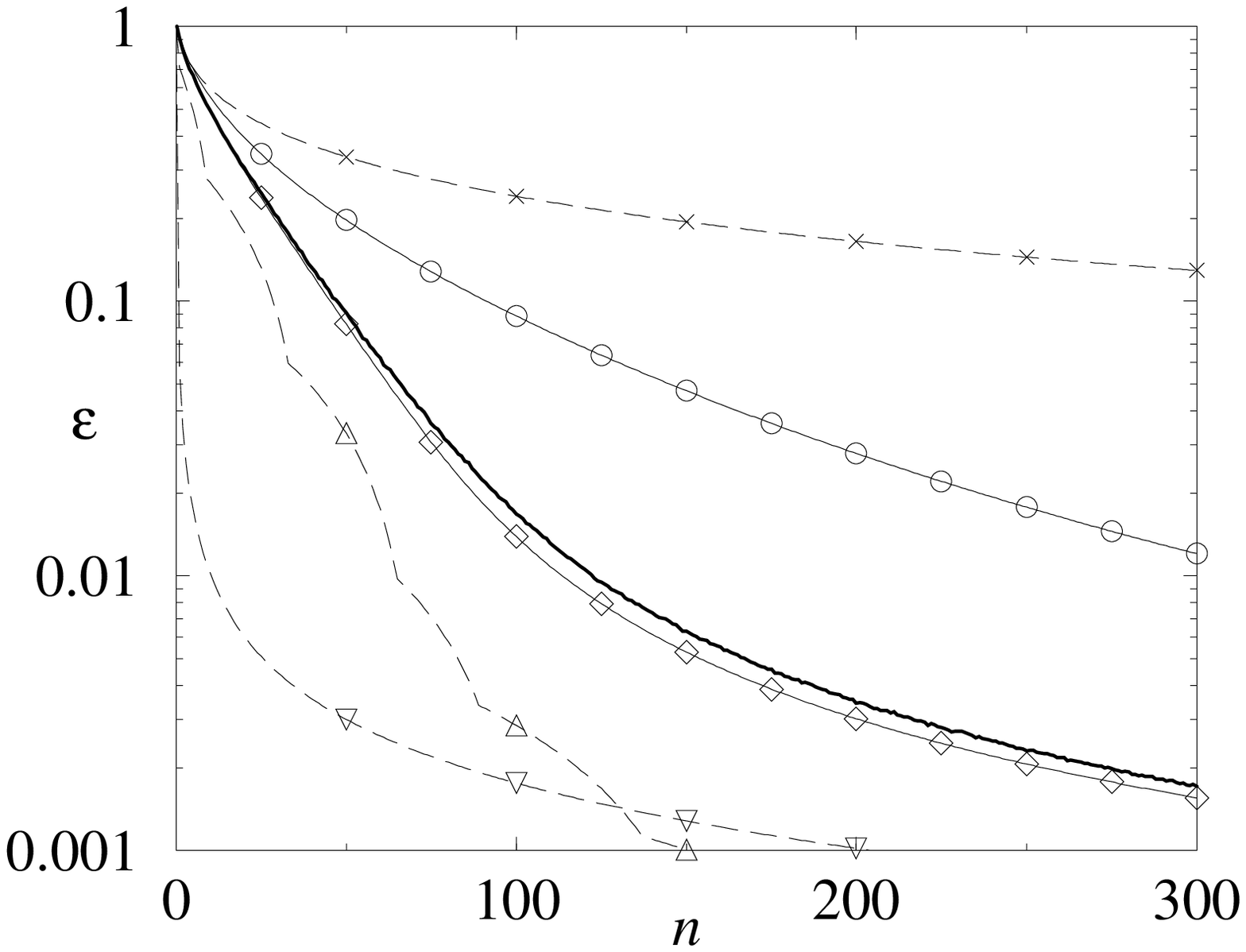, height=5.5cm}
\end{center}
\caption{Learning curve for a GP with RBF covariance function and
inputs uniformly drawn from $x\in[0,1]^d$, for dimension $d=4$ and
length scale $l=0.3$. Left: noise level $\noise=0.05$. Right:
$\noise=0.001$. Note that, for lower $\noise$, the OV bound becomes
looser as expected (it approaches zero for $\noise\to 0$).
\label{fig:RBFcube}
}
\end{figure}
Figure~\ref{fig:RBFcube} shows corresponding results in higher
dimension ($d=4$), at two different noise levels $\noise$. One
observes in particular that the OV lower bound becomes looser as
$\noise$ decreases; this is as expected since for $\noise\to 0$ the
bound actually becomes void ($\epsOV\to 0$). The Plaskota bound also
appears to get looser for lower $\noise$, though not as
dramatically. (Note that the kinks in the Plaskota curve are not an
artifact: For larger $d$ the multiplicities of the different eigenvalues
can be quite large; the value of $\epsPl$ can become dominated by
one such block of degenerate eigenvalues, and kinks occur where the
dominant block changes.) The TWO upper bound, finally, is
only weakly affected by the value of $\noise$ and quite loose throughout.

\begin{figure}
\begin{center}
\epsfig{file=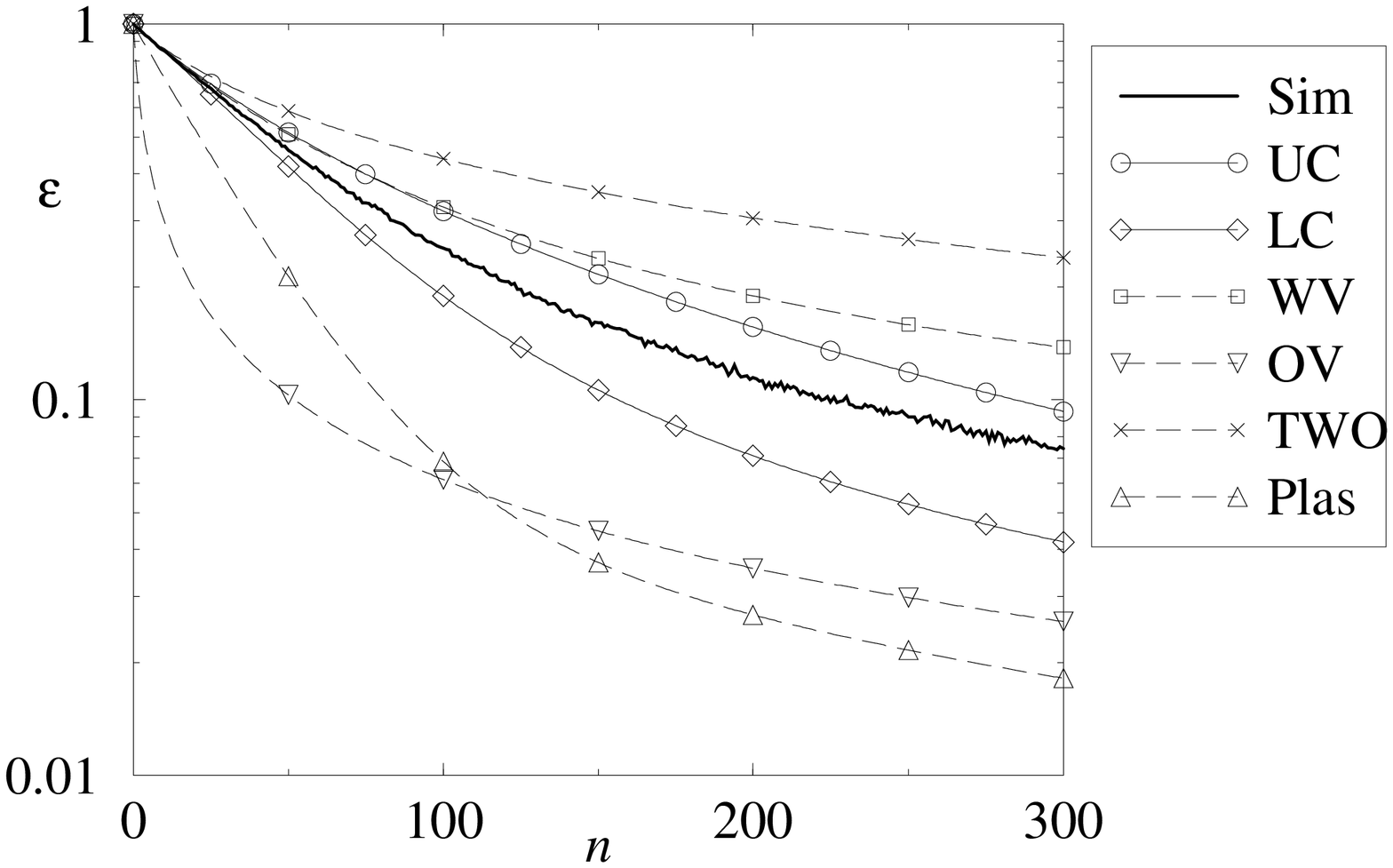, height=5.5cm}
\end{center}
\caption{Learning curve for a GP with RBF covariance function (length
scale $l=0.01$) and inputs drawn from a Gaussian distribution in $d=1$
dimension; noise level $\noise=0.05$. Note that the LC approximation
provides less of a good representation of the overall shape of the
learning curve here than for the previous examples with uniform input
distributions. The curve labelled WV shows our generalized version of
the Williams-Vivarelli bound (see eq.~(\protect\ref{epsWVnonunif})).
\label{fig:RBFGaussian_oned}
}
\end{figure}
\begin{figure}
\begin{center}
\epsfig{file=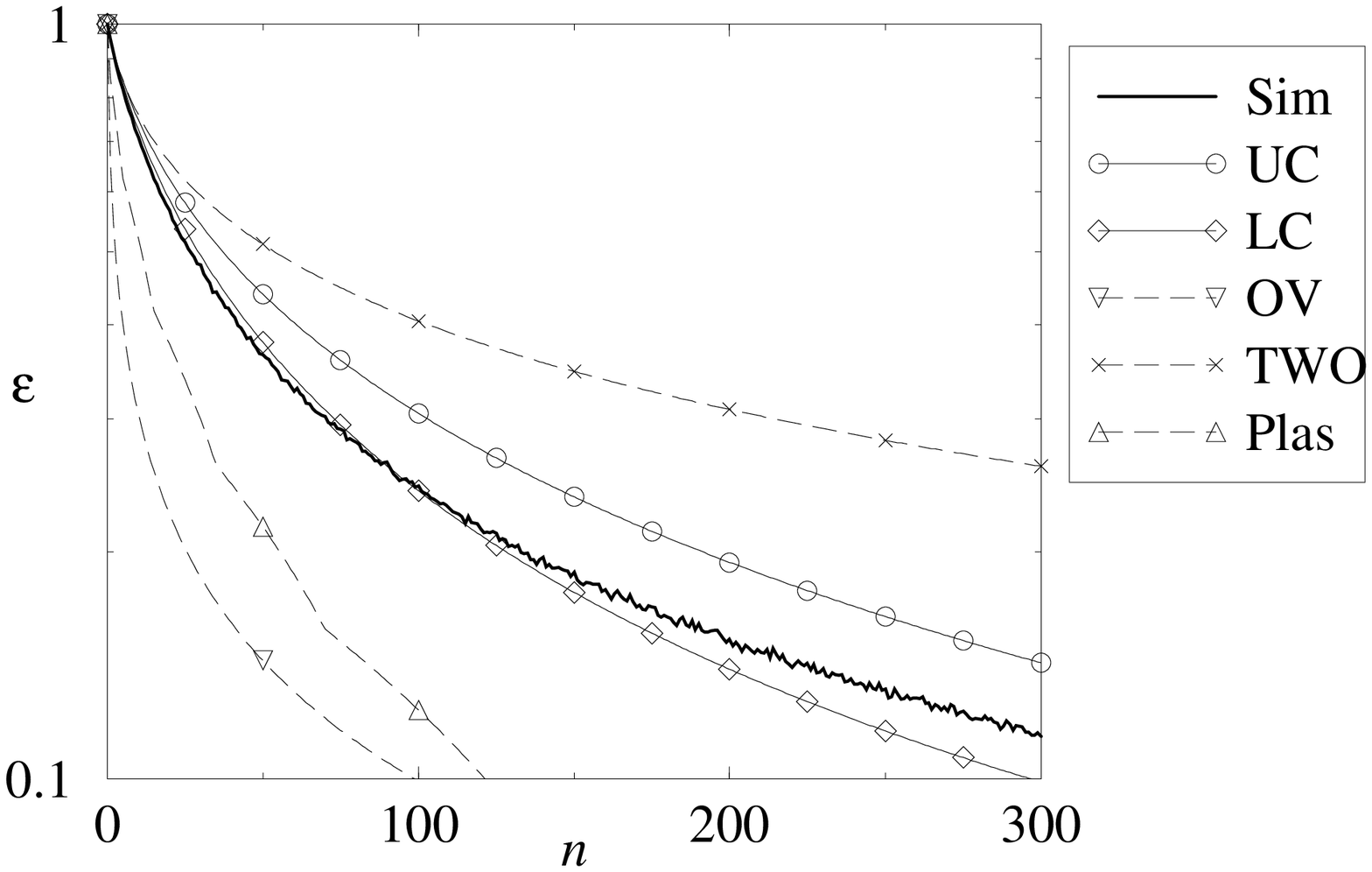, height=5.5cm}%
\hspace*{0.05cm}%
\epsfig{file=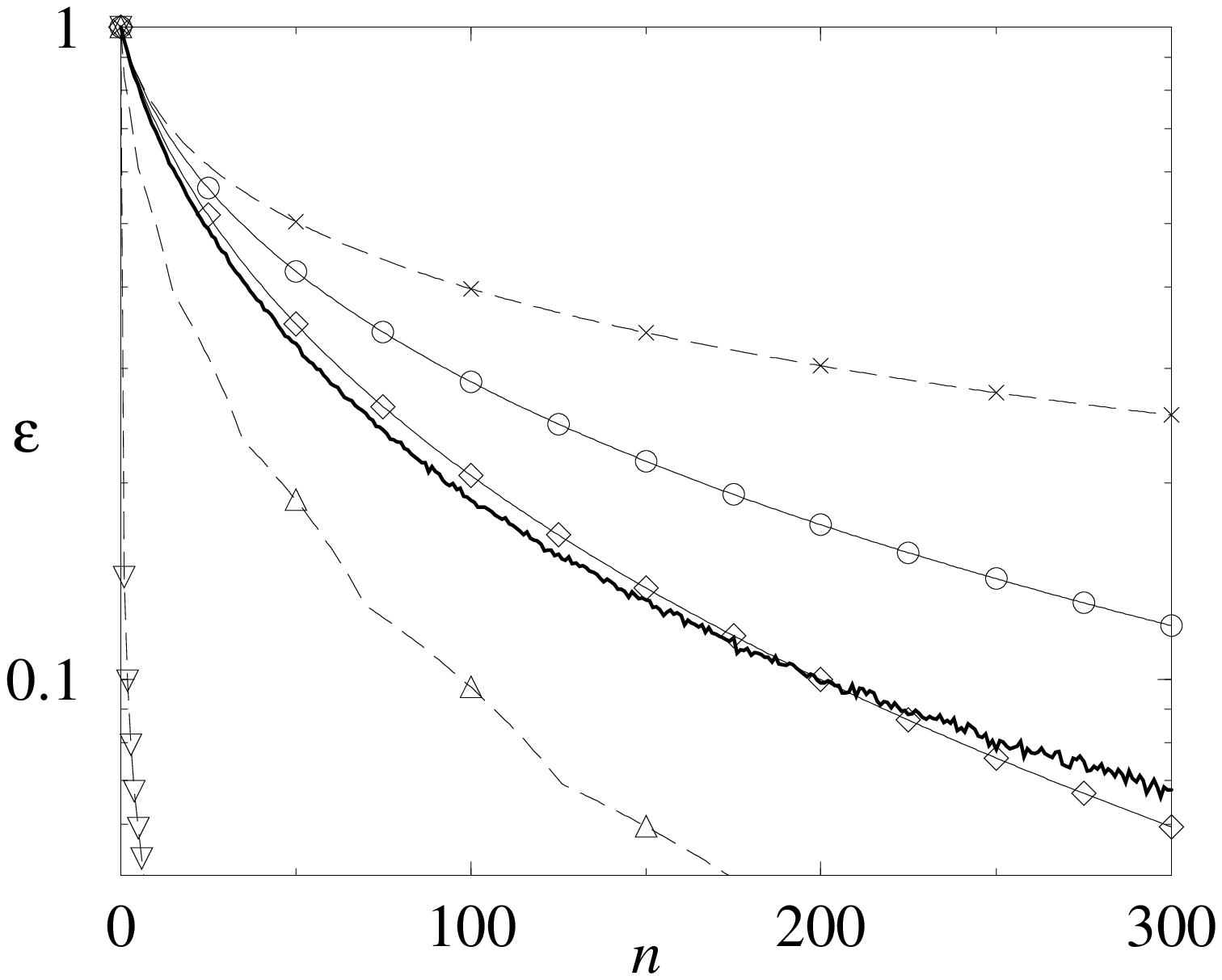, height=5.5cm}
\end{center}
\caption{Learning curve for a GP with RBF covariance function (length
scale $l=0.3$) and inputs drawn from a Gaussian distribution in $d=4$
dimensions. Left: noise level $\noise=0.05$. Right:
$\noise=0.001$. The OV lower bound is very loose for this smaller
noise level. Note that, in contrast to all previous examples, there is
a range of $n$ here where the true learning curve lies {\em below}
the LC approximation.
\label{fig:RBFGaussian}
}
\end{figure}
All results shown so far pertain to uniform input distributions (over
$[0,1]^d$). We now move to the last of our three scenarios, a GP with
an RBF covariance function and inputs drawn from a Gaussian
distribution (see App.~\ref{app:spectra} for details). In
Figure~\ref{fig:RBFGaussian_oned} we see that in $d=1$ the
(generalized) WV bound is still reasonably tight, while the LC
approximation now provides less of a good representation of the
overall shape of the learning curve than for the case of uniform input
distributions. However, as in all previous examples, the LC and UC
approximations still bracket the true learning curve (and come closer
to it than the bounds). One is thus lead to speculate whether the
approximations we have derived are actually
bounds. Figure~\ref{fig:RBFGaussian} shows this not to be the case,
however: In $d=4$, the true learning curve drops visibly below the LC
approximation in the small $n$ regime, and so the latter cannot be a
lower bound. The low noise case ($\noise=0.001$) shown here
illustrates once more that the OV lower bound ceases to be useful for
small noise levels.

In summary, of the approximations which we have derived, the LC
approximation performs best. While we know on theoretical grounds that
it will be accurate for large noise levels $\noise$, the examples
shown above demonstrate that it produces predictions close to the true
learning curves even for the more realistic case of low noise levels
(compared to the prior variance). As a general trend, agreement
appears to be better for the case of uniform input distributions.

It is interesting at this stage to make a connection to the recent
work of Malzahn and Opper~\cite{MalOpp01}. They devised an elegant way
of approaching the learning curve problem from the point of view of
statistical physics, calculating the relevant partition function
(which is an average over data sets) using a so-called Gaussian
variational approximation. The result they find for the Bayes error is
{\em identical} to the LC approximation under the condition that
$\eps(x) = \lav \eps(x,D) \rav_D$, the $x$-dependent generalization
error averaged over all data sets, is independent of $x$. Otherwise,
they find a result of the same functional form, $\eps(n) =
\tr(\LL^{-1} + \eta\mident)^{-1}$, but the self-consistency equation
for $\eta$ is more complicated than the simple relation
$\eta=n/(\eps+\noise)$ obtained from the LC
approximation\eq{epsLC}. The LC approximation would thus be expected
to perform less well for such ``non-uniform'' scenarios. This agrees
qualitatively with our above findings: For the scenario with a
Gaussian input distribution, the LC approximation is of poorer quality
than for the cases with uniform input distributions%
\footnote{
It is easy to see that in these cases $\eps(x)$ is indeed independent
of $x$; the absence of effects from the boundaries of the hypercube
comes from the periodic boundary conditions that we are using.
}
over $[0,1]^d$.

\iffalse
Several observations can be made from Figure~\ref{fig:comparison}. (1)
Lower bounds: The MW bound is not tight, as expected. The OV bound
performs well asymptotically ($\eps\ll\noise$), but is rather loose
initially, when $\eps\gg\noise$, in particular for smaller noise
levels.  (2) Upper bounds: The UO bound gives a good representation of
the overall shape of the learning curve only in the asymptotic regime
(most clearly visible for the SE covariance function), and is
generally rather loose. The TWO bound is fairly insensitive to
$\noise$ (compare graphs a, b to c,d) and tight only for small $n$.
The WV bound (available only in $d=1$) works well for the OU
covariance function, but less so for the SE case. As expected, it is
not useful in the asymptotic regime because it always remains above
$\noise$.  (3) In all the examples, the true learning curve lies
between the UC and LC curves. In fact we would {\em conjecture} (but
have no proof) that these two approximations provide upper and lower
bounds on the learning curves, at least for stationary covariance
functions; they certainly lie within the region allowed by existing
bounds. (4) Finally, the LC approximation comes out as the clear
winner: For $\noise=0.1$ (Figure~\ref{fig:comparison}c,d), it is
indistinguishable from the true learning curves. But even in the other
cases it represents the overall shape of the learning curves very
well, both for small $n$ and in the asymptotic regime; the largest
deviations occur in the crossover region between these two regimes.
\fi

\section{Improving the approximations}
\label{sec:improving}

In the previous section we saw that in our test scenarios the LC
approximation\eq{epsLC} generally provides the closest theoretical
approximation to the true learning curves. This may appear somewhat
surprising, given that we made two rather drastic approximations in
deriving\eq{epsLC}: We treated the number of training examples $n$ as
a continuous variable, and we decoupled the average of the right hand
side of~\(G_update) into separate averages over numerator and
denominator. We now investigate whether the LC prediction for the
learning curves can be further improved by removing these
approximations.

We begin with the effect of $n$, the number of training examples,
taking only discrete (rather than continuous) values.  Starting
from\eq{G_update}, averaging numerator and denominator separately as
before and introducing the auxiliary variable $v$ as
in\eqq{vdef}{Gsquared}, we obtain
\be
\eps(n+1)-\eps(n) = \frac{1}{\noise+\eps}\frac{\partial\eps}{\partial v}
\label{LCimpr_eqn}
\ee
instead of\eq{LC_eqn}. It is possible to interpolate between the two
equations by writing
\be
\frac{1}{\delta}[\eps(n+\delta)-\eps(n)] =
\frac{1}{\noise+\eps}\frac{\partial\eps}{\partial v}
\label{LC_delta}
\ee
Then $\delta=1$ corresponds to\eq{LCimpr_eqn}, which is the equation
we wish to solve (discrete $n$), while in the limit $\delta\to 0$ we
retrieve\eq{LC_eqn}. To proceed, we treat $\delta$ as a perturbation
parameter and assume that the solution of\eq{LC_delta} can be
expanded as
\[
\eps = \eps_0 + \delta \eps_1 + \order(\delta^2)
\]
where $\eps_0 \equiv \epsLC$. Expanding both sides of\eq{LC_delta} to
first order in $\delta$ yields
\[
\frac{\partial \eps_0}{\partial n} + \delta 
\frac{\partial \eps_1}{\partial n} + \half\delta \frac{\partial^2
\eps_0}{\partial n^2} + \order(\delta^2) = \left(\frac{1}{\noise+\eps_0}
- \delta\,\frac{\eps_1}{(\noise+\eps_0)^2}\right)\left(
\frac{\partial\eps_0}{\partial v} + \delta\,
\frac{\partial\eps_1}{\partial v}\right)
\]
Comparing the coefficients of the $\order(\delta^0)$ terms gives us
back\eq{LC_eqn} for $\eps_0$, while from the $\order(\delta)$ terms we
get
\be
\frac{\partial \eps_1}{\partial n} - 
\frac{1}{\noise+\eps_0} \frac{\partial\eps_1}{\partial v} 
= -\half \frac{\partial^2 \eps_0}{\partial n^2}
-\frac{\eps_1}{(\noise+\eps_0)^2}\frac{\partial\eps_0}{\partial v}
\label{LCo_eqn}
\ee
This can again be solved using characteristics (see
App.~\ref{app:char}), with the result
\be
\eps_1 = (\noise+\eps_0) \frac{n(a_2^2-a_3)}{(1-na_2)^2}, \qquad
a_k = (\noise+\eps_0)^{-k} \ 
\tr\left(\LL^{-1}+\frac{n}{\noise+\eps_0}\mident\right)^{-k}
\label{eps_one}
\ee
Setting $\delta$ back to $1$ to have the case of discrete $n$
in\eq{LC_delta}, we then have
\[
\epsLCo = \eps_0 + \eps_1 \equiv \epsLC + \eps_1
\]
as the improved LC approximation that takes into account the effects
of discrete $n$ (up to linear order in a perturbation expansion in
$\delta$). We see that the correction term\eq{eps_one} is zero at
$n=0$ as it must since $\epsLC$ gives the exact result $\eps=\tr\LL$
there. It can also be shown that $\eps_1<0$ for all nonzero $n$. This
can be understood as follows: The decrease of $\eps(n)$ becomes
smaller (in absolute value) as $n$ increases. Comparing\eq{LC_eqn}
and\eq{LCimpr_eqn}, we see that the continuous $n$-approximation
effectively averages the decrease term over the range $n\ldots n+1$
rather than evaluating it at $n$ itself; it therefore produces a
smaller decrease in $\eps(n)$. The true decrease for discrete $n$ is
larger, and so one expects the correction $\eps_1$ to be negative, in
agreement with our calculation.

\begin{figure}
\begin{center}
\epsfig{file=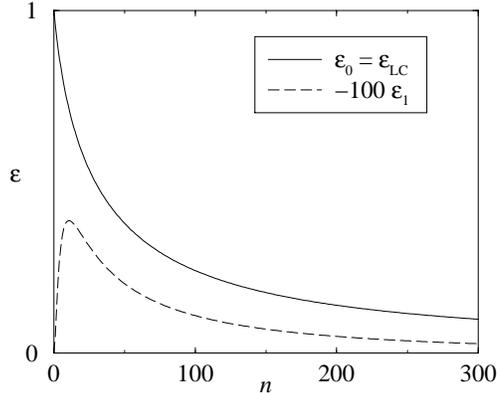, height=5.5cm}
\end{center}
\caption{Solid line: LC approximation $\epsLC$ for the learning curve
of a GP with RBF covariance function and Gaussian inputs. Parameters
are as in Figure~\protect\ref{fig:RBFGaussian}(left): Dimension $d=4$,
length scale $l=0.3$, noise level $\noise=0.05$. Dashed line: First
order correction $\eps_1$ arising from the discrete nature of the
number of training examples $n$. Note that $\eps_1$ has been
multiplied by -100 to make it positive and visible on the scale of the
values of $\epsLC$.
\label{fig:eps_one}
}
\end{figure}
In Figure~\ref{fig:eps_one} we show $\epsLC$ and $\eps_1$ for one of
the scenarios considered earlier; the results are typical also of what
we find for other cases.  The most striking observation is the
smallness of $\eps_1$: Its absolute value is of the order of 1\% of
$\epsLC$ or less, and consequently $\epsLC$ and $\epsLCo$ are
indistinguishable on the scale of the plot. On the one hand, this is
encouraging: Given that $\eps_1$ is already so small, one would expect
higher orders in a perturbation expansion in $\delta$ to yield even
smaller corrections. Thus $\epsLCo$ is likely to be very close to the
result that one would find if the discrete nature of $n$ was taken
into account exactly. On the other hand, we also conclude that
treating $n$ as discrete is {\em not} sufficient to turn the LC
approximation into a lower bound on the learning curve; in
Fig~\ref{fig:RBFGaussian_oned}, for example, the curve for $\epsLCo$
would lie essentially on top of the one for $\epsLC$ and so still be
significantly above the true learning curve for small $n$.

It is clear at this stage that in order to improve the LC
approximation significantly one would have to address the decoupling
of the numerator and denominator averages in\eq{G_update}. Generally,
if $a$ and $b$ are random variables, one can evaluate the average of
their ratio perturbatively as
\[
\lav\frac{a}{b}\rav = 
\lav \frac{\lav a\rav + \Delta a}{\lav b\rav + \Delta b} \rav = 
\frac{\lav a \rav}{\lav b \rav} + \frac{\lav a\rav}{\lav
b\rav^3}\lav(\Delta b)^2\rav - \frac{1}{\lav b \rav^2} \lav \Delta a
\Delta b\rav + \ldots
\]
up to second order in the fluctuations. (This idea was used
in~\cite{Sollich94f} to calculate finite $N$-corrections to the
$N\to\infty$ limit of a linear learning problem.) To apply this to the
average of the right hand side of\eq{G_update} over the new training
input $x_{n+1}$ and all previous ones, one would set
\[
a = \Gfluc(n) \new\new\T\Gfluc(n), \qquad
b = \noise+\new\T\Gfluc(n)\new
\]
One then sees that averages such as $\lav a b\rav$, required in $\lav
\Delta a \Delta b\rav = \lav ab\rav - \lav a \rav \lav b \rav$,
involve fourth-order averages $\lav \phi_i(x) \phi_j(x) \phi_k(x)
\phi_l(x) \rav_x$ of the components of $\new$. In contrast to the
second-order averages $\lav \phi_i(x)\phi_j(x)\rav = \delta_{ij}$,
such fourth-order statistics of the eigenfunctions do not have a
simple, covariance function-independent form. Even if they were known,
however, one would end up with averages over the entries of the matrix
$\Gfluc$ which cannot be reduced to $\eps = \tr\lav\Gfluc \rav$ (\eg\
by derivatives with respect to auxiliary parameters). Separate
equations for the change of these averages with $n$ would then be
required, generating new averages and eventually an infinite hierarchy
which cannot be closed. We thus conclude that a perturbative approach
is of little use in improving the LC approximation beyond the
decoupling of averages. The approach of~\cite{MalOpp01} thus looks
more hopeful as far as the derivation of systematic corrections to the
approximation is concerned.

\section{How good can bounds and approximations be?}
\label{sec:limits}

In this final section we ask whether there are limits of principle on
the quality of theoretical predictions (either bounds or
approximations) for GP learning curves.  Of course this question is
meaningless unless we specify what information the theoretical curves
are allowed to exploit. Guided by the insight that all predictions
discussed above depend (at least for uniform covariance
functions) on the eigenvalues of the covariance function only (and of
course the noise level $\noise$), we ask: How tight can bounds and
approximations be if they use only this eigenvalue spectrum as input?

To answer this question, it is useful to have a simple scenario with
an arbitrary eigenvalue spectrum for which learning curves can be
calculated exactly. Consider the case where the input space consists
of $N$ discrete points $\xa$; the input distribution is arbitrary,
$P(\xa) = \pa$ with $\sum_\al \pa = 1$. Take the covariance
function to be degenerate in the sense that there are no correlations
between different points: $C(\xa,\xb) = \ca \delta_{\al\bet}$.
The eigenvalue equation\eq{eigenfunc} then becomes simply
\[
\sum_\bet C(\xa,\xb) \phi(\xb) \pb = \ca \pa \phi(\xa) = \lam
\phi(\xa)
\]
so that the $N$ different eigenvalues are $\lam_\al = \ca\pa$. The
eigenfunctions are $\phi_\al(\xb) = \pa^{-1/2} \delta_{\al\bet}$,
where the prefactor follows from the normalization condition
$\sum_\gam \pg\phi_\al(\xg)\phi_\bet(\xg) = \delta_{\al\bet}$. Note that
by choosing the $\ca$ and $\pa$ appropriately, the $\lam_\al$ can be
adjusted to any desired value in this setup. The same still holds even
if we require the covariance function to be uniform, \ie, $\ca$ to be
independent of $\al$.

A set of $n$ training inputs is, in this scenario, fully characterized
by how often it contains each of the possible inputs $\xa$; we call
these numbers $\na$. The generalization error is easy to work out
from\eq{basic_result}, using 
\[
(\Ph\T\Ph)_{\al\bet} = \sum_\gam n_\gam \phi_\al(\xg)\phi_\bet(\xg) =
(\na/\pa) \delta_{\al\bet}.
\]
This shows that $\Ph\T\Ph$ is a diagonal matrix, and thus
from\eq{basic_result}
\be
\eps(D) = \tr (\LL^{-1}+\noiseinv\Ph\T\Ph)^{-1}
= \sum_\al \left(\lam_\al^{-1} + \noiseinv \na/\pa\right)^{-1}
= \sum_\al \lam_\al \frac{\noise}{\noise + \na\lam_\al/\pa}
\label{epsD_deg}
\ee
This has the expected form: The contribution of each eigenvalue is
reduced according to the ratio of the noise level $\noise$ and the
signal $\na\lam_\al/\pa=\na\ca$ received at the corresponding input
point. To average this over all training sets of size $n$, one notices
that $\na$ has a binomial distribution, so that
\[
\eps(n) = \sum_\al \lam_\al \sum_{\na=0}^n \colvec{n}{\na} \pa^{\na}
(1-\pa)^{n-\na} \frac{\noise}{\noise+\na\lam_\al/\pa}
\]
Writing $(\noise + \na\lam_\al/\pa)^{-1} = \noiseinv\int_0^1\! dr\
r^{\na\lam_\al/\pa\noise}$, we can perform the sum over $\na$ and
obtain
\be
\eps(n) = \int_0^1 \! dr\ \sum_\al \lam_\al
\left(1-\pa+\pa r^{\lam_\al/\pa\noise}\right)^n
\label{exact_deg}
\ee
as the final result; the integral over $r$ can easily be performed
numerically for given set of eigenvalues $\lam_\al$ and input
probabilities $\pa$. Note that, having done the calculation for a
finite number $N$ of discrete inputs points (and therefore of
eigenvalues), we can now also take $N$ to infinity and therefore
analyse scenarios (such as the ones studied in
Sec.~\ref{sec:comparison}) with an infinite number of nonzero
eigenvalues.

A simple limiting case will now tell us about the quality of
eigenvalue-dependent upper bounds on learning curves. Assume that one
of the $p_\al$ is close to $1$, whereas all the other ones are close
to $0$. From\eq{exact_deg}, one then sees that only the contribution
from the eigenvalue $\lam_\al$ with $\pa\approx 1$ is reduced as $n$
increases%
\footnote{This holds if $n$ is not too large, more precisely if
$n\pb\ll 1$ for all the `small' $\pb$ ($\bet\neq \al$).}
while all other ones remain unaffected, so that
\be
\eps(n) \approx \tr\LL - \lam_\al + \frac{\lam_\al\noise}{\noise +
n\lam_\al}
=
\tr\LL - \frac{n\lam_\al^2}{\noise + n\lam_\al} \geq \tr\LL - \lam_\al
\label{eps_limiting}
\ee
If $\lam_\al\ll\tr\LL$, then we can make the reduction in
generalization error arbitrarily small. It thus follows that there is
no non-trivial upper bound on learning curves that takes only the
eigenvalue spectrum of the covariance function as input. [Accordingly,
the two non-asymptotic upper bounds (WV and TWO) discussed in
Sec.~\ref{sec:comparison} both contain other information, via the
weighted averages of $C_s^2(x)=C^2(0,x)$ in\eq{epsWV} and the
average of $C(x,x)\phi_i^2(x)$ in\eq{epsTWO_ci}.]  In particular, this
implies that our UC approximation cannot be an upper bound (even
though the results for all scenarios investigated above suggested that
it might be). Furthermore, our result shows that lower bounds on the
generalization error (\eg\ the OV or Plaskota bounds) can be
arbitrarily loose. A similar observation holds for {\em upper} bounds
on the (data set-averaged) {\em training} error $\eps_{\rm t}$: Opper
and Vivarelli showed~\cite{OppViv99} that their $\epsOV$ actually also
provides an upper bound for this quantity (so that the two errors
`sandwich' the bound, $\eps_{\rm t}\leq \epsOV \leq \eps$). In the
present case it is easy to calculate $\eps_{\rm t}$ explicitly; we
omit details and just quote the result
\[
\eps_{\rm t} \approx \frac{\lam_\al\noise}{\noise + n\lam_\al} \leq \lam_\al
\]
By taking $\lam_\al\ll\tr\LL$, one then sees that the upper bound
$\epsOV$ can be made arbitrarily loose for any fixed $n$ (and that the
ratio of training and generalization error can be made arbitrarily
small).

\begin{figure}
\begin{center}
\epsfig{file=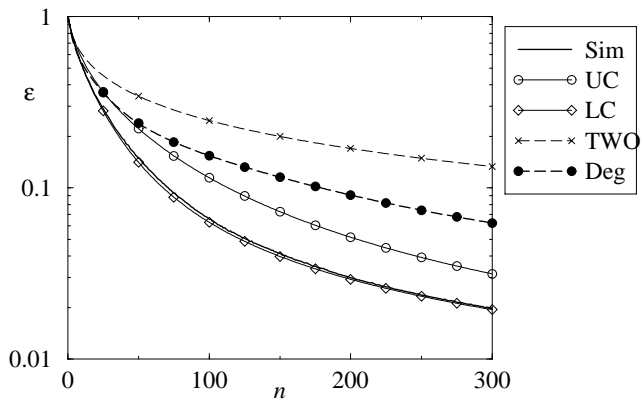, height=5.5cm}
\end{center}
\caption{Learning curve for a GP with RBF covariance function and
inputs uniformly drawn from $x\in[0,1]^d$. Parameters are as in
Figure~\protect\ref{fig:RBFcube}(left): Dimension $d=4$, length scale
$l=0.3$ and noise level $\noise=0.05$. The curves Sim (true learning
curve, from numerical simulations), UC, LC and TWO are also as in
Figure~\protect\ref{fig:RBFcube}(left). The curve labelled ``Deg''
shows the exact learning curve for the degenerate scenario (outputs
for different inputs are uncorrelated) with {\em exactly the same
spectrum of eigenvalues} $\lam_i$ of the covariance function (and
uniform prior variance $C(x,x)$). The curves Sim and Deg differ
significantly, showing that learning curves cannot be predicted
reliably based on eigenvalues alone.
\label{fig:degenerate}
}
\end{figure}

One may object that the above limit of most of the $\pa$ tending to
zero is unrealistic because it implies that the corresponding prior
variances $\ca = \lam_\al/\pa$ would become very large. Let us
therefore now restrict the prior variance to be uniform, $\ca=c$. It
then follows that $\lam_\al = c/\pa$ and hence $\pa =
\lam_\al/\tr\LL$. With this assumption, only the $\lam_\al$ and
$\noise$ remain as parameters affecting the learning
curve\eq{exact_deg}. The results for an eigenvalue spectrum from one
of the situations covered in Sec.~\ref{sec:comparison} are shown in
Figure~\ref{fig:degenerate}. The main conclusion to be drawn is that
the learning curves for the present scenario are quite different from
the ones we found earlier, even though the eigenvalue spectra and
noise levels are, by construction, precisely identical. This
demonstrates that theoretical predictions for learning curves which
only take into account the eigenvalue spectrum of a covariance
function cannot universally match the true learning curves with a high
degree of accuracy; the quality of approximation will vary depending
on details of the covariance function and input distribution that are
not encoded in the spectrum. Note that Figure~\ref{fig:degenerate}
also provides a concrete example for the fact that the UC
approximation is not in general an upper bound on the true learning
curve; in fact, it here {\em underestimates} the true $\eps(n)$ quite
significantly.

\begin{figure}
\begin{center}
\epsfig{file=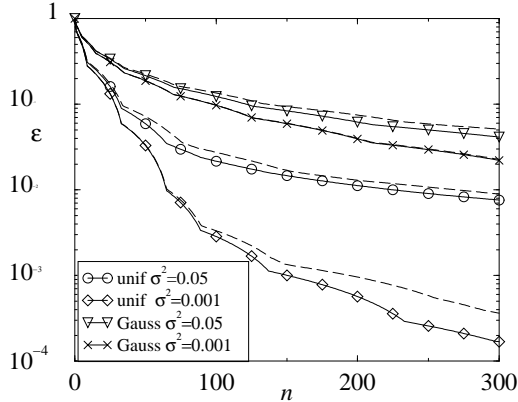, height=5.5cm}
\end{center}
\caption{Comparison of the Plaskota bound (solid lines) and the lowest
generalization error achievable for single data sets of size $n$
within the degenerate scenario (dashed lines). The eigenvalue spectra
used to construct the curves are those for an RBF covariance function
with length scale $l=0.3$, in $d=4$ dimensions, and for the input
distributions (uniform over $[0,1]^d$, or Gaussian) shown in the
legend; the noise level $\noise$ is also given there. Note that for a
given $n$, the curves become closer for lower $\noise$; this is as
expected since for $\noise\to 0$ the Plaskota bound can be saturated
for a specific data set (see text).
\label{fig:Plaskota_limit}
}
\end{figure}
We can also use the present scenario to assess whether, as a bound on
the generalization error resulting from a {\em single} training set,
the Plaskota bound\eq{Plaskota} could be significantly improved. We
focus on the case of uniform covariance functions, where\eq{epsD_deg}
becomes
\be
\eps(D) = \sum_\al \lam_\al \frac{\noise}{\noise + \na \tr\LL}
\label{epsD_uniform_deg}
\ee
For any assignment of the $\na$ there is at least one training set of
size $n=\sum_\al \na$ for which the generalization error is given
by\eq{epsD_uniform_deg}. Minimizing numerically over the $\na$ for
each given $n$, we find the curves shown in
Figure~\ref{fig:Plaskota_limit}, where the Plaskota bounds for the
same eigenvalue spectra are also shown. The curves are quite close to
each other, implying that the Plaskota bound cannot be significantly
tightened as a single data set bound (assuming, as throughout, that
the improved bound would again only be based on the covariance
function's eigenvalue spectrum). In the limit $\noise\to 0$, the
bound---which then reduces to the MW bound~\eq{epsMW}---cannot be
tightened at all, as setting $\na=1$ for $\al=1\ldots n$ and $\na=0$
for $\al\geq n+1$ in\eq{epsD_uniform_deg} shows.

Within the simple degenerate scenario introduced in this section, we
finally comment briefly on a universal relation recently proposed by
Malzahn and Opper~\cite{MalOpp01}. They suggest considering an
empirical estimate of the (Bayes) generalization error, which is
obtained by replacing the average over all inputs $x$ by one over the
$n$ training inputs $x_i$:
\[
\epsemp(D) = \frac{1}{n} \sum_{i=1}^n \eps(x_i,D)
\]
Within the approximations of their calculation, the data set average
of this quantity is then universally linked to a modified version of the
true generalization error:
\be
\lav\epsemp(D)\rav_D = \lav
\frac{\eps(x)}{\noise+\eps(x)}\rav_x
\label{universal}
\ee
Note that the average over data sets is on the `inside' of the
fraction on the right hand side, through the definition of
$\eps(x)=\lav \eps(x,D)\rav_D$. Within our degenerate scenario, we can
calculate both sides of\eq{universal} explicitly, but find no obvious
relation between the two sides. However, if we move the data set
average on the right hand side to the outside, we do (after a brief
calculation, the details of which we omit) find a simple result:
\be
\lav\epsemp(D_{n+1})\rav_{D_{n+1}} = \lav \lav
\frac{\eps(x,D_n)}{\noise+\eps(x,D_n)}\rav_{\!\!x} \, \rav_{\!\!D_n}
\label{mod_universal}
\ee
As indicated by the subscripts, the left hand side of this relation is
to be evaluated for data sets of size $n+1$ rather than $n$. The
result\eq{mod_universal} is remarkable in that it holds for any
eigenvalue spectrum and any input distribution (within the degenerate
scenario considered here). We take this as a hopeful sign that some
universal link between true and empirical generalization errors, along
the lines derived by~\cite{MalOpp01} within their approximation, may
indeed exist.

\section{Conclusion}

In summary, we have derived an exact representation of the average
generalization $\eps$ error of Gaussian processes used for regression,
in terms of the eigenvalue decomposition of the covariance
function. Starting from this, we obtained two different approximations
(LC and UC) to the learning curve $\eps(n)$. Both become exact in the
large noise limit; in practice, one generically expects the opposite
case ($\noise/C(x,x)\ll 1$), but comparison with simulation results
shows that even in this regime the new approximations perform
well.

The LC approximation in particular represents the overall shape of the
learning curves very well, both for `rough' (OU) and `smooth' (RBF)
Gaussian priors, and for small as well as for large numbers of
training examples $n$. It is not perfect, but does generally get
substantially closer to the true learning curves than existing bounds
(two of which, due to Plaskota and to Williams and Vivarelli, we
generalized to a wider range of scenarios). For situations with
non-uniform input distributions the predictions of the LC
approximation tend to be less accurate, and we linked this observation
to recent work by Malzahn and Opper~\cite{MalOpp01} on the effects of
non-uniformity across input space. Their result, which reduces to the
LC approximation for sufficiently uniform scenarios, may in other
cases provide better approximations, but this has to be traded off
against the higher computational cost that would be involved in
actually evaluating the predictions.

We then discussed how the LC approximation could be improved. The
effects of discrete $n$ can be incorporated to leading order, but were
seen to be relatively minor; on the other hand, the second
approximation involved in the derivation (decoupling of averages)
appears difficult to improve on within our framework.

Finally, we investigated a simple ``degenerate'' Gaussian process
learning scenario, where the outputs corresponding to different inputs
are uncorrelated. This provided us with a means of assessing whether
there are limits on the quality of approximations and bounds that take
into account only the eigenvalue spectrum of the covariance
function. We found indeed that such limits exist: There can be no
nontrivial upper bound on the learning curve of this form, and
approximations are necessarily of limited quality because different
covariance functions with the same eigenvalue spectrum can produce
rather different learning curves. We also found that as a bound on the
generalization error for {\em single} data sets (rather than its
average over data sets) the Plaskota bound is close to being tight.
Whether a tight lower bound on the {\em average} learning curve exists
remains an open question; one plausible candidate worth investigating
would be the average generalization error of our degenerate scenario,
minimized over all possible input distributions for a fixed eigenvalue
spectrum.

There are a number of open problems. One is whether a sub-class of GP
learning scenarios can be defined for which the covariance function's
eigenvalue spectrum is sufficient to predict the learning curves
accurately. Alternatively, one could ask what (minimal) extra
information beyond the eigenvalue spectrum needs to be taken into
account to arrive at accurate learning curves for all possible GP
regression problems. Finally, one may wonder whether the eigenvalue
decomposition we have chosen, which explicitly depends on the input
distribution, is really the optimal one. On the one hand, recent work
(see \eg~\cite{WilSee00}) appears to answer this question in the
affirmative. On the other hand, the variability of learning curves
among GP covariance functions with the same eigenvalue spectrum
suggests that the eigenvalues alone do not provide sufficient
information for accurate predictions. One may therefore speculate that
eigendecompositions with respect to other input distributions (\eg,
maximum entropy ones) might not suffer from this problem. We leave
these challenges for future work.

%\subsubsection*{Acknowledgements} 
{\bf Acknowledgements:} We would like to thank Chris Williams, Manfred
Opper and D\"orte Malzahn for stimulating discussions, and the Royal
Society for financial support through a Dorothy Hodgkin Research
Fellowship.

\appendix

\section{Solving for the LC approximation}
\label{app:char}

In this appendix we describe how to solve eqns.\eqq{LC_eqn}{LCo_eqn}
for the LC approximation and its first order correction, using the
method of characteristic curves.  The method applies to partial
differential equations of the form $a\, \partial f/\partial x +
b\,\partial f/\partial y = c$, where $f=f(x,y)$ and $a,b,c$ can be
arbitrary functions of $x,y,f$. Viewing the solution as a surface in
$x,y,f$-space, one can then show~\cite{John78} that if the point
$(x_0,y_0,f_0)$ belongs to the solution surface then so does the
entire characteristic curve $(x(t),y(t),f(t))$ defined by
\[
\frac{dx}{dt} = a, \qquad 
\frac{dy}{dt} = b, \qquad 
\frac{df}{dt} = c, \qquad (x(0),y(0),f(0)) = (x_0,y_0,f_0)
\]
The solution surface can then be recovered by combining an
appropriate one-dimensional family of characteristic curves.

Denote the generalization error predicted by the LC approximation as
$\eps_0(n,v)$, with $v$ the auxiliary parameter introduced
in\eqq{vdef}{Gsquared}. It is the solution of the equation\eq{LC_eqn}
\[
\frac{\partial\eps_0}{\partial n} -
\frac{1}{\noise+\eps_0}\frac{\partial\eps_0}{\partial v} = 0
\]
subject to the initial conditions $\eps_0(n=0,v)= \tr(\LL^{-1} +
v\mident)^{-1}$. These give us a family of solution points which the
characteristic curves have to pass through, namely $(n(0)=0, v(0)=v_0,
\eps_0(0) = \tr(\LL^{-1}+v_0\mident)^{-1})$. The equations for the
characteristic curves are 
\[
\frac{dn}{dt} = 1, \qquad
\frac{dv}{dt} = -\frac{1}{\noise+\eps_0}, \qquad
\frac{d\eps_0}{dt} = 0
\]
and can be integrated to give 
\be
n=n(0)+t = t, \qquad
v=v(0)-\frac{t}{\noise+\eps_0} = v_0 - \frac{t}{\noise+\eps_0}, 
\qquad
\eps_0 = \eps_0(0) = \tr(\LL^{-1}+v_0\mident)^{-1}
\label{epso_chars}
\ee
Eliminating $t$ (the curve parameter) and $v_0$ (which parameterizes
the family of initial points) gives the required solution $\eps_0 =
\tr\{\LL^{-1}+ [v+n/(\noise+\eps_0)]\mident\}^{-1}$. The LC
approximation\eq{epsLC} is obtained by setting $v=0$.

For the first order correction $\eps_1$, we have to solve the
equation\eq{LCo_eqn}
\[
\frac{\partial \eps_1}{\partial n} - 
\frac{1}{\noise+\eps_0} \frac{\partial\eps_1}{\partial v} 
= -\half \frac{\partial^2 \eps_0}{\partial n^2}
-\frac{\eps_1}{(\noise+\eps_0)^2}\frac{\partial\eps_0}{\partial v}
\]
with the initial condition (explained in the main text)
$\eps_1(n=0,v)=0$. Hence a suitable family of initial solution points
is $(n(0)=0,v(0)=v_0, \eps_1(0)=0)$. The characteristic curves must
obey
\[
\frac{dn}{dt} = 0, \qquad
\frac{dv}{dt} = -\frac{1}{\noise+\eps_0}, \qquad
\frac{d\eps_1}{dt} = -\half \frac{\partial^2 \eps_0}{\partial n^2}
-\frac{\eps_1}{(\noise+\eps_0)^2}\frac{\partial\eps_0}{\partial v}
\]
The solutions for the $n(t)$ and $v(t)$-dependence are therefore the same
as before, and as a result $\eps_0$ is again constant along the
characteristic curves. For the derivatives of $\eps_0$ that appear,
one finds after some algebra
\[
\frac{\partial \eps_0}{\partial v} = -(\noise+\eps_0)^2\frac{a_2}{1-n
a_2}, \qquad
\frac{\partial^2\eps_0}{\partial n^2} = 
-2(\noise+\eps_0)\frac{a_2^2-a_3}{(1-n a_2)^3}
\]
Here we have used the definitions from\eq{eps_one}; because both $a_2$
and $a_3$ only depend on $n$ and $v$ in the combination
$v+n/(\noise+\eps_0)$, they are also constant along the characteristic
curves. Using also that $n=t$ from\eq{epso_chars}, the equation for
$\eps_1$ becomes
\[
\frac{d\eps_1}{dt} = (\noise+\eps_0)\frac{a_2^2-a_3}{(1-t a_2)^3}
 + \eps_1 \frac{a_2}{1-t a_2}
\]
This linear differential equation is easily integrated; using the
initial condition $\eps_1(0) = 0$ one finds
\[
\eps_1 = (\noise+\eps_0) \frac{t(a_2^2-a_3)}{(1-ta_2)^2}
\]
Eliminating $t$ again via $t=n$ finally gives the
solution\eq{eps_one}.

\section{The Plaskota bound and its extension}
\label{app:Plaskota}

To summarize the proof of Plaskota's lower bound on learning
curves~\cite{Plaskota90} and generalize it to non-uniform covariance
functions, we find it most convenient to work with a discrete space of
inputs $\xa$, $\al=1\ldots N$, with input probabilities $P(\xa) =
\pa$. (The case where inputs can vary continuously can be approximated
arbitrarily well by such a scenario, either by discretizing the input
space on a sufficiently fine grid or by taking the $\xa$ as a large
number $N$ of samples from the input distribution and setting
$\pa=1/N$.) A function $\theta$ on these inputs can then be thought of
as a vector $\thv$ with elements $\theta(\xa)$; similarly, the
covariance function $C(\xa,\xb)\equiv C_{\al\bet}$ becomes an $N\times
N$ matrix $\CC = \lav \thv \thv\T \rav$. The eigenvalue
equation\eq{eigenfunc} and the corresponding normalization condition
then read
\[
\sum_\bet C_{\al\bet}\pb\phi_i(\xb) = \lam_i \phi_i(\xa), \qquad
\sum_\al \phi_i(\xa)\phi_j(\xa)\pa = \delta_{ij}
\]
We can write these in a more compact form: If we set
$\LL=\mbox{diag}(\lam_1\ldots \lam_N)$ and $\PP=\mbox{diag}(p_1\ldots
p_N)$, denote $\phiv_i$ the vector with entries $\phi_i(\xa)$ and
$\Phii$ the $N\times N$ matrix whose columns are the $\phiv_i$, we
have
\[
\CC\PP\Phii = \Phii\LL, \qquad \Phii\T\PP\Phii = \mident
\]
and hence
\[
\CC = \Phii\LL\Phii\T
\]
One can also write this in the form of a more conventional matrix
diagonalization:
\[
\PP^{1/2}\CC\PP^{1/2} = (\PP^{1/2}\Phii)\LL(\PP^{1/2}\Phii)\T, \qquad
(\PP^{1/2}\Phii)\T(\PP^{1/2}\Phii) = \mident
\]
It follows that the $\lam_i$ are the eigenvalues of the
matrix $\PP^{1/2}\CC\PP^{1/2}$ and that
\be
\lav C(x,x)\rav_x \equiv \tr \PP\CC = \tr \PP^{1/2}\CC\PP^{1/2} = \tr \LL
\label{Cav_discr}
\ee
in agreement with\eq{Cav}.

Plaskota now considers the case of {\em generalized} training sets
consisting of $n$ generalized observations $\yl = \Lv_l\T\thv +
\eta_l$. Each $\yl$ is a linear combinations of the values of the
function $\theta(x)$, with coefficients specified by the vector
$\Lv_l$, corrupted by additive Gaussian noise $\eta_l$ which as before
we take to be of variance $\noise$. The conventional scenario, where
each training point contains a (corrupted) observation of $\theta(x)$
at a single point $x=x^{\al(l)}$, corresponds to $\Lv_l =
\ev_{\al(l)}$. Here we denote $\ev_\al$ the $\al$-th unit vector (with
the $\al$-th component equal to one and all others zero) and write the
$l$-th training input as $x^{\al(l)}$, with $\al(l) \in \{1\ldots N\}$.

After the observation of the $\yl$, we will have some posterior
covariances $\lav \Delta\theta(\xa) \Delta\theta(\xb)
\rav$. Collecting these into an $N\times N$ matrix $\Cov$ one can
write them explicitly as
\be
\Cov = \CC - \CC\Lv(\Lv\T\CC\Lv + \noise \mident)^{-1} \Lv\T \CC =
(\CC^{-1} + \noiseinv \Lv \Lv\T)^{-1}
\label{cov_Plaskota}
\ee
where $\Lv$ is an $N\times n$ matrix whose columns are the $\Lv_l$.
The first version of this result can be seen as a generalization
of\eq{basic_inf_b}; the second version is the corresponding
generalization%
\footnote{%
Note that the matrices $\Cov$ and $\Gfluc$ are related in the same way
as $\CC$ and $\LL$; explicitly, one has $\Cov=\Phii\Gfluc\Phii\T$.
}%
\ of\eq{basic_result} in a different guise. It then follows that the
generalization error is given by
\be
\eps(D) = \sum_\al \pa\lav (\Delta \theta(\xa))^2 \rav =
\tr \PP\Cov = \tr\PP\CC -
\tr[\Lv\T\CC\PP\CC\Lv(\MM + \noise\mident)^{-1}]
\label{Plaskota_eps}
\ee
where we have defined the $n\times n$ matrix $\MM = \Lv\T\CC\Lv$. Its
trace (which will become important shortly) is $\tr\MM = \sum_l
\Lv_l \CC \Lv_l$; for a conventional training set with training inputs
$x^{\al(l)}$ this becomes $\tr\MM = \sum_l \ev_{\al(l)}\T\CC
\ev_{\al(l)} = \sum_l C(x^{\al(l)},x^{\al(l)})$.

So far everything is exact. The idea behind Plaskota's bound is now as
follows: For any given {\em conventional} training set with $n$ training
inputs, minimize $\eps(D)$ over all {\em generalized} training sets of size
$n$ which have the same value of $\tr\MM$. The result then clearly
gives a lower bound on $\eps(D)$ for the given training set.

To carry out this minimization, one can proceed as follows. The matrix
$\MM$ is symmetric and can be diagonalized, so that $\MM = \OO \etam
\OO\T$ where $\OO$ is an orthogonal $n\times n$ matrix and $\etam$ is
a diagonal matrix whose entries $\eta_i$ are the eigenvalues of
$\MM$. Since $\MM$ is positive semi-definite, the $\eta_i$ are
non-negative%
\footnote{
We actually assume, for simplicity of presentation, that all the
$\eta_i$ are nonzero. Equation~(\protect\ref{eps_intermed}) still
holds if some $\eta_i$ vanish, but the derivation is then slightly
more complicated: One has to replace $\etam$ by a smaller diagonal
matrix which contains only the positive eigenvalues and $\OO$ by the
matrix whose columns are the corresponding eigenvectors of $\MM$.
}%
; we also assume them to be ordered in descending order, $\eta_1\geq
\ldots \geq \eta_N$. It then follows that $(\MM+\noise\mident)^{-1} =
\OO(\etam + \noise\mident)^{-1} \OO\T$, so that
\[
\eps(D) = \tr\LL -
\tr[\OO\T\Lv\T\CC\PP\CC\Lv\OO(\etam + \noise\mident)^{-1}]
\]
where we have also used that $\tr\PP\CC = \tr\LL$ from\eq{Cav_discr}.
Now, if we define the $N\times n$ matrix $\QQ =
\CC^{1/2} \Lv \OO \etam^{-1/2}$ 
and write $\Ct = \CC^{1/2}\PP\CC^{1/2}$, then
$\etam^{1/2}\QQ\T\Ct\QQ\etam^{1/2} = \OO\T \Lv\T \CC\PP\CC\Lv\OO$ and so
\be
\eps(D) = \tr\LL - 
\tr[\QQ\T\Ct\QQ \etam(\etam + \noise\mident)^{-1}] = 
\tr\LL - \sum_{i=1}^n (\QQ\T\Ct\QQ)_{ii} \frac{\eta_i}{\eta_i + \noise}
\label{eps_intermed}
\ee
To minimize $\eps(D)$, we need to make the sum over $i$ maximal.  If
we write the $n$ columns of $\QQ$ as vectors $\QQ_i$ then
$(\QQ\T\Ct\QQ)_{ii} = \QQ_i\T\Ct\QQ_i$; also, from the definition of
$\QQ$ we have $\QQ\T\QQ = \mident$, implying that the $\QQ_i$ are
orthonormal. Now it is easy to see that $\Ct = \CC^{1/2}\PP\CC^{1/2}$
%
%since $|\Ct-\lambda\mident|$ $=$
%$|\CC^{1/2}\PP\CC^{1/2}-\lambda\mident|$ $=$ 
%$|\PP^{1/2}\CC\PP^{1/2}-\lambda\mident|$ the matrix $\Ct$
%
has the same eigenvalues as $\PP^{1/2}\CC\PP^{1/2}$, namely, the
$\lam_i$. Assuming these to be ordered in a descending sequence, the
largest value that $\QQ_i\T\Ct\QQ_i$ can take is therefore
$\lam_1$. Because of the orthonormality of the $\QQ_i$, one has
similarly that $\sum_{i\leq k} \QQ_i\T\Ct\QQ_i \leq \sum_{i\leq k}
\lam_i$ for all $k\leq n$. Since the terms $\eta_i/(\eta_i+\noise)$
form a descending sequence (due to our assumed ordering of the
$\eta_i$), it is then clear that the optimal situation is the one
where $\QQ_1\T\Ct\QQ_1$, multiplying the largest value
$\eta_1/(\eta_1+\noise)$, has the largest possible value $\lam_1$,
$\QQ_2\T\Ct\QQ_2$ has the largest possible value subject to this,
which is $\lam_2$, and so on. Plaskota indeed proved formally that
\[
\eps(D) \geq \tr\LL - \sum_{i=1}^n \frac{\lam_i\eta_i}{\eta_i + \noise}
\]
Finally, one can now optimize over the $\eta_i$, subject to the constraint
that we imposed initially, \ie, that $\tr\MM = \sum_i \eta_i$ equals
the sum $S = \sum_l C(x^{\al(l)},x^{\al(l)})$. This gives the final
Plaskota bound
\be
\eps(D) \geq \tr\LL - \max_{\{\eta_i\}} 
\sum_{i=1}^n \frac{\lam_i\eta_i}{\eta_i + \noise} = 
\min_{\{\eta_i\}} \left(\sum_{i=1}^n \frac{\lam_i\noise}{\eta_i +
\noise}\right) + \sum_{i\geq n+1} \lam_i
\label{Plaskota_app}
\ee
as given in\eq{Plaskota}. Plaskota also proved this bound to be
realizable by an appropriate choice of the generalized observations
$\Lv_l$. His original derivation only applied to uniform covariance
functions, but the above proof sketch shows that this restriction is
not in fact necessary.

To evaluate the Plaskota bound in practice it is desirable to have a
more explicit expression for the minimum over the $\eta_i$.
Introducing a Lagrange multiplier $\alpha$ for the constraint $\sum_i
\eta_i = S$, one finds by differentiation of\eq{Plaskota_app} that
each positive $\eta_i$ must satisfy
\be
- \frac{\lam_i\noise}{(\eta_i + \noise)^2} + \alpha = 0
\label{opt_cond}
\ee
while for the vanishing $\eta_i$ the quantity on the left hand side
has to be positive, \ie, $\lam_i \leq \alpha\noise$. Due to the
ordering of the $\lam_i$, one sees from this that at the minimum the
first $k$ of the $\eta_i$ will be nonzero and the rest will be zero,
with $k\leq n$ a number to be determined. Assume we know $k$. Then
from\eq{opt_cond} the nonzero $\eta_i$ are given by
\[
\eta_i = \left(\frac{\lam_i\noise}{\alpha}\right)^{1/2} - \noise
\]
Using that $\sum_{i\leq k} \eta_i = S$, one then finds $\alpha^{-1/2} =
(S+k\noise)/(\sigma \sum_{j \leq k} \lam_j^{1/2})$ and hence
\be
\eta_i = \lam_i^{1/2} \frac{S+k\sigma^2}{\sum_{j\leq k} \lam_j^{1/2}}
- \noise
\label{etas}
\ee
In order for the value of $k$ that we assumed to be the correct one,
this expression needs to be positive for $i=1\ldots k$, while for
$i=k+1\ldots n$ it must be zero or negative (as follows from
$\lam_i\leq \alpha\noise$). Due to the ordering of the $\lam_i$, it
is sufficient to check whether\eq{etas} gives $\eta_k>0$ and
$\eta_{k+1}\leq 0$. The $k$ that satisfies these conditions%
\footnote{
There is a unique $k$ with this property; this follows because the
minimum in~(\protect\ref{Plaskota_app}) is over a convex function of
the $\eta_i$ and therefore unique.
}
gives the minimum in\eq{Plaskota_app}; using\eq{etas}, the bound can
then be simplified to
\[
\eps(D) \geq \frac{\noise}{S+k\noise}
\left({\sum_{i\leq k} \lam_i^{1/2}}\right)^2 + \sum_{i\geq k+1} \lam_i
\]
In the example scenarios for which we evaluated the bound, we found
that $k=n$ in the vast majority of cases. A sensible numerical
procedure is therefore to check first whether for $k=n$
equation\eq{etas} gives $\eta_n>0$. If yes, then $k=n$ gives the
required optimum; if no, then $k$ should be decreased until the
conditions $\eta_k>0$ and $\eta_{k+1}\leq 0$ are met.

\section{Eigenvalue spectra for the example scenarios}
\label{app:spectra}

We explain first how the covariance functions with periodic boundary
conditions for $x\in[0,1]^d$ are constructed. Consider first the case
$d=1$. The periodic RBF covariance function is defined as
\be
C(x,x') = \sum_{r=-\infty}^\infty c(x-x'-r)
\label{periodic}
\ee
where $c(x-x')=\exp[-|x-x'|^2/(2l^2)]$ is the original covariance
function; for the periodic OU case we use instead $c(x-x') =
\exp(-|x-x'|/l)$. One sees that for sufficiently small $l$ ($l\ll 1$),
only the $r=0$ term makes a significant contribution, except when $x$
and $x'$ are within $\approx l$ of opposite ends of the input space
(so that either $x-x'+1$ or $x-x'-1$ are of order $l$). We therefore
expect the periodic covariance functions and the conventional
non-periodic ones to yield very similar learning curves, as long as
the length scale of the covariance function is smaller than the size of
the input domain.

The advantage of having a periodic covariance function is that its
eigenfunctions are simple Fourier waves and the eigenvalues can be
calculated by Fourier transformation. This can be seen as follows. For
the assumed uniform input distribution on $[0,1]$, the defining
equation for an eigenfunction $\phi(x)$ with eigenvalue $\lam$ is,
from\eq{eigenfunc}
\[
\lav C(x,x') \phi(x') \rav_{x'} = \int_0^1 \! dx' \ C(x,x')\phi(x') = 
\lam \phi(x).
\]
Inserting\eq{periodic} and assuming that $\phi(x)$ is continued
periodically outside the interval $[0,1]$, this becomes
\[
\sum_{r=-\infty}^\infty \int_0^1\! dx' \ c(x-x'-r) \phi(x') = 
\sum_{r=-\infty}^\infty \int_r^{r+1}\! dx' \ c(x-x') \phi(x'-r) =
\int_{-\infty}^\infty \! dx' \ c(x-x') \phi(x') = \lam \phi(x)
\]
It is well known that the solutions of this eigenfunction equation are
Fourier waves $\phi_q(x) = e^{2\pi iqx}$ for integer (positive or
negative) $q$, with corresponding eigenvalues
\[
\lam_q = \int_{-\infty}^\infty \!dx\ c(x) e^{-2 \pi i q x}
\]
The eigenvalues $\lam_q$ are real since $c(x)=c(-x)$ (this follows
from the requirement that the covariance function $C(x,x')$ must be
symmetric). The eigenfunctions for $q\neq 0$ are complex in the form
given but can be made explicitly real by linearly transforming
the pair $\phi_q(x)$ and $\phi_{-q}(x)$ into $(1/\sqrt{2})\cos(2\pi
qx)$ and $(1/\sqrt{2})\sin(2\pi qx)$.

All of the above generalizes immediately to higher input dimension
$d$. One defines
\[
C(x,x') = \sum_{r} c(x-x'-r)
\]
where $r$ now runs over all $d$-dimensional vectors with integer
components; the argument $x-x'-r$ of $c(\cdot)$ is now a
$d$-dimensional real vector. The eigenvalues of this periodic
covariance function are then given by
\[
\lam_q = \int\!dx\ c(x) e^{-2 \pi i q\cdot x}
\]
They are indexed by $d$-dimensional integer vectors $q$; the
integration is over all real-valued $d$-dimensional vectors $x$, and
$q\cdot x$ is the conventional dot product between
vectors. Explicitly, one derives that for the periodic RBF covariance
function
\[
\lam_q = (2\pi)^{d/2} l^d e^{-(2\pi l)^2||q||^2/2}
\]
For the periodic OU covariance function, on the other hand, one has
\be
\lam_q = \kappa_d l^d [1+(2\pi l)^2 ||q||^2]^{-(d+1)/2}
\label{OU_eigenvals}
\ee
with $\kappa_d=2$, $2\pi$, $8\pi$ for $d=1,2,3$, respectively; for
general $d$, $\kappa_d = \pi^{(d-1)/2} 2^d \Gamma((d+1)/2)$ where
$\Gamma(z) = (z-1)!$ is the Gamma function.

All the bounds and approximations in principle require traces over the
whole eigenvalue spectrum, corresponding to sums over an infinite
number of terms. Numerically, we perform the sums over all eigenvalues
up to some suitably large maximal value $q_{\rm max}$ of $||q||$. The
remaining small eigenvalue tail of the spectrum is then treated by
approximating $||q||$ as a continuous variable $\tilde{q}$ and
integrating over it from ${q}_{\rm max}$ to infinity, with the
appropriate weighting for the number of vectors $q$ in a shell
$\tilde{q} \leq ||q|| \leq \tilde{q} + d\tilde{q}$. To check that this
procedure gave accurate results, we always verified that the
numerically calculated $\tr\LL$ agreed with the known value of
$C(x,x)$.

The third scenario we consider is that of a conventional RBF kernel
$C(x,x') = \exp[-||x-x'||^2/(2l^2)]$ with a nonuniform input
distribution which we assume to be an isotropic zero mean Gaussian,
$P(x)$ $\propto$ $\exp[-||x||^2/(2\sigma_x^2)]$. The eigenfunctions
and eigenvalues are worked out in~\cite{ZhuWilRohMor98} for the case
$d=1$; the eigenvalues are labelled by a non-negative integer $q$ and
given by
\[
\lam_q = (1-\beta) \beta^q
\]
where
\[
\beta^{-1} = 1 + r/2 + \sqrt{r^2/4+r}, 
%
%\alpha = \sqrt{r\beta} = 1-\beta, 
%
\qquad
r = l^2/\sigma_x^2.
\]
As expected, only the ratio $r$ of the length scales $l$ and
$\sigma_x$ enters since the overall scale of the inputs is immaterial.
(To avoid this trivial invariance we fixed $\sigma^2_x=1/12$; this
specific value gives the same variance for each component
of $x$ as for the uniform distribution over $[0,1]^d$ used in the
other scenarios.) For $d>1$, this result generalizes immediately
because both the covariance function and the input distribution
factorize over the different input
components~\cite{OppViv99,WilSee01}. The eigenvalues are therefore
just products of the eigenvalues for each component, and indexed by a
$d$-dimensional vector $q$ of non-negative integers:
\be
\lam_q = (1-\beta)^d \beta^s, \qquad s = \sum_{i=1}^d q_i
\label{nonunif_lambdas}
\ee
One sees that the eigenvalues will come in blocks: All vectors $q$
with the same $s=\sum_i q_i$ give the same $\lam_q$. Numerically, we
therefore only have to store the different eigenvalues and their
multiplicities, which can be shown to be%
\footnote{%
There is a nice combinatorial way of seeing this. Imagine a row of
$d+s+1$ holes. Each hole is either empty or contains one ball, except
the holes at the two ends which have one ball placed in them
permanently. Now consider $d-1$ identical balls distributed across the
$d+s-1$ free holes, and let $q_i$ ($i=1\ldots d$) be the number of
free holes between balls $i-1$ and $i$ (where balls $0$ and $d$ are
the fixed balls in the holes at the left and right end,
respectively). The $q_i$ are non-negative integers, and their sum
equals the number of free holes, giving $\sum_i q_i=(d+s-1)-(d-1) =
s$. The number of different assignments of the $q_i$ is thus identical
to the number of different arrangements of $d-1$ identical balls across
$d+s-1$ holes, hence the result.
}
$(d+s-1)!/[s!(d-1)!]$. With this trick so many eigenvalues can be
treated by direct summation that a separate treatment of the neglected
eigenvalues (via an integral, as above) is unnecessary.

%\bibliography{/home/psollich/neural_nets/references,local}
%\bibliographystyle{unsrt}

\end{document}